# Momentum-Transfer Framework Unifies High-Velocity Impact and Failure Across Materials, Geometries, and Scales

Yasara Dharmadasa[a,1], Nicholas Jaegersberg[a,1], Ara Kim[a], Jizhe Cai[b] and Ramathasan Thevamaran[a,*]

[a]Department of Mechanical Engineering, University of Wisconsin–Madison, Madison, WI, USA
[b]Department of Industrial & Manufacturing Engineering, FAMU–FSU College of Engineering, Tallahassee, FL, USA

ABSTRACT

Materials that dissipate energy efficiently under high-speed impacts—from micrometeoroid strikes on spacecraft to ballistic penetration in protective systems—are essential for maintaining structural integrity in extreme environments. Yet, predicting and comparing impact performance across materials and geometries remains challenging because analytical models require knowledge of the material response under the relevant impact conditions, while empirical descriptions remain tied to the regimes in which they are calibrated. These limitations become particularly important at the extreme deformation rates accessed in microprojectile impact experiments. Here, we directly examine the momentum and energy transferred from the projectile during impact. Microprojectile impact experiments spanning varied projectile diameters, target thicknesses, and impact velocities, supported by targeted macroscale tests and a broad survey of independent impact studies, reveal a consistent behavior across materials, scales, projectile geometries, and impact conditions: momentum transfer reaches its maximum at the ballistic limit. The resulting momentum bound identifies the ballistic limit as the maximum momentum that a target can sustain without perforation—its momentum capacity—and limits the maximum projectile velocity reduction during perforation to the ballistic-limit velocity. The observed momentum-transfer trends further capture the influence of material cohesion, failure, and target inertia on the perforation response. We also demonstrate that specific energy absorption and specific momentum transfer can produce contradictory rankings of ballistic performance because the specific-energy representation remains influenced by the projectile–target mass ratio. Specific momentum transfer and specific momentum capacity provide a more consistent basis for comparing projectile–target configurations and tracking changes in ballistic response across scale. These findings establish momentum transfer as a unifying framework for interpreting projectile perforation and provide a mechanics-based foundation for developing improved analytical models and protective systems.

## 1. Introduction

Impact protection spans an extraordinary range of length scales and loading conditions, from conventional ballistic threats encountered by military personnel and first responders to micron- and sub-millimeter-scale particles striking spacecraft at hypervelocity[1–4]. Developing enabling material technologies to withstand these extreme environments has been a challenge due to multiple requirements that often present a trade-off, for example, achieving high energy absorption, strength, and stiffness at ultra-lightweight for affording protections while simultaneously enabling mobility, functionality, and reduced payload[5, 6]. Advances in lightweight, high-performance composites have sought to address these competing requirements by providing alternatives to bulky metallic and ceramic armor systems. For instance, low-density fibrous composites based on carbon, aramid, and ultra-high-molecular-weight polyethylene fibers can provide high energy absorption at a fraction of the weight, while retarding failure through multiscale deformation mechanisms[7–10]. This emphasis on lightweight protection has made mass-normalized energy metrics, particularly specific energy absorption, a widely used measure for comparing ballistic performance.

*Corresponding author.
ORCID(s): 0000-0002-3240-2296 (Y. Dharmadasa); 0009-0006-7636-2827 (N. Jaegersberg); 0000-0002-5776-8921 (A. Kim); 0000-0003-2803-2111 (J. Cai); 0000-0001-5058-6167 (R. Thevamaran)

[1]These authors contributed equally to this work.

Recent studies on micron-thick nanostructured materials, such as polymers[11–14], nanofibrous mats[15–18], and nanolattices[19–21], have reported specific energy absorption, $E_a^* = E_a/m_{\text{plug}}$, values that are an order of magnitude higher than those of state-of-the-art bulk protective materials (see Fig. 1(a)), often attributed to nanoscale size effects and favorable mesoscale interactive morphologies. Here, $E_a = \frac{1}{2}m_p(v_i^2 - v_r^2)$ represents the kinetic energy lost by the projectile during impact, and $m_{\text{plug}}$ denotes the target mass within the projected area of the projectile. These emerging materials are typically tested using a miniaturized ballistic testing platform, the laser-induced projectile impact test (LIPIT)[22] (Fig. 1(b)), where small volumes of nanostructured thin-film targets ($\sim$ 200 $\mu$m $\times$ 200 $\mu$m $\times$ $h$, with $h$ ranging from hundreds of nanometers to tens of micrometers) are impacted by 3–30 $\mu$m-diameter projectiles at velocities ranging from approximately 100 m/s to 1 km/s. In contrast, conventional macroscale ballistic experiments typically employ millimeter-scale projectiles and targets ranging from millimeters to centimeters in thickness, as illustrated by the gas-gun setup (Fig. 1(c)). This remarkable increase in specific energy absorption for sub-micron thick targets raises a fundamental question: *what drives this improvement as the projectile–target system is reduced in length scale?* Understanding the origin of this enhanced performance could help guide the design of next-generation lightweight protection systems.

To better understand the formulation of specific energy absorption, consider a spherical projectile of diameter $D$ impacting a flat target of thickness $h$. The specific energy absorption term simplifies to

$$E_a^* = \frac{1}{3}\frac{\rho_p}{\rho_t}\frac{D}{h}\left(v_i^2 - v_r^2\right), \tag{1}$$

where $\rho_p$ and $\rho_t$ are the projectile and target densities, and $v_i$ and $v_r$ are the impact and residual projectile velocities, respectively. This expression separates the explicit geometric and density dependencies through the ratios $D/h$ and $\rho_p/\rho_t$. Yet, experiments maintaining comparable geometric and density ratios exhibit higher $E_a^*$ at smaller length scales [23], reflecting changes in material response, strain rate, deformation and failure mechanisms, and potential experimental scale effects that are captured through the $v_i$–$v_r$ relationship. Understanding how these mechanisms alter projectile deceleration is therefore central to interpreting the measured energy absorption.

The $v_i$–$v_r$ response has been extensively studied across projectile–target materials and geometries, leading to numerous analytical and empirical descriptions. A commonly used generalized representation is

$$v_r = a\left(v_i^p - v_{bl}^p\right)^{1/p}, \tag{2}$$

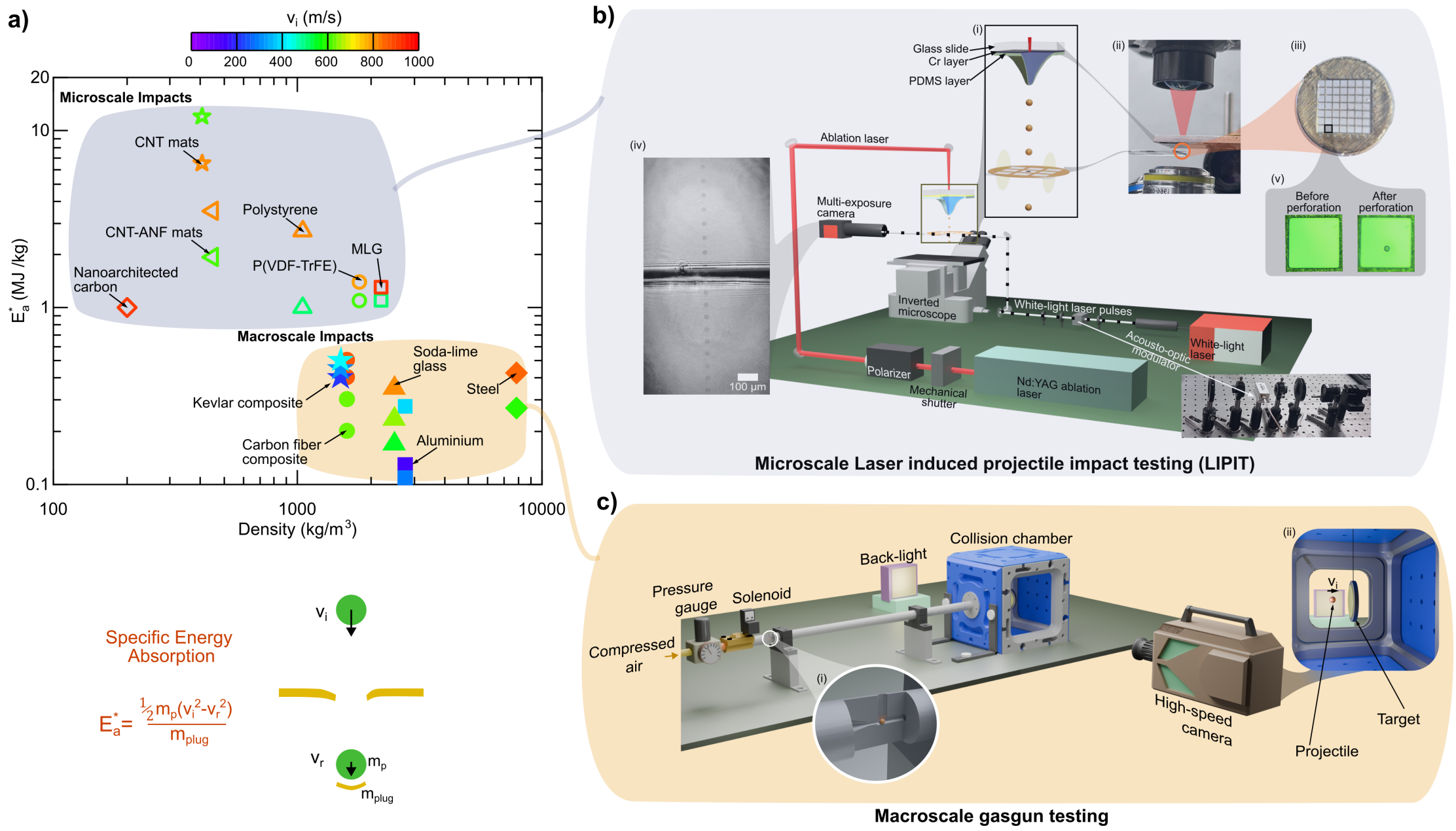

**Figure 1: Impact performance across scales and experimental platforms.** (a) Reported specific energy absorption for different material systems tested at micro- and macroscales, showing markedly higher values in microscale impact experiments. (b) Laser-induced projectile impact test (LIPIT) used for microscale impact experiments. A pulsed Nd:YAG laser ablates the chromium layer of the launch pad, generating rapidly expanding plasma that drives the PDMS layer and launches microparticles toward polystyrene films supported on TEM grids; insets (i)–(iii) show the launch and target configuration. A sequence of white-light laser pulses enables multi-exposure imaging of the projectile trajectory recorded in inset (iv), and inset (v) shows an enlarged cell before and after perforation. (c) Macroscale gas-gun impact setup. The projectile is placed within the steel launch tube shown in inset (i), and accelerated using compressed air pulses. A high-speed camera records the projectile motion before and after impact as illustrated in inset (v).

where $v_{bl}$ is the ballistic-limit velocity. This relation originates from the Recht–Ipson formulation [24], for which the commonly used simplified form corresponds to $a = 1$ and $p = 2$. The relation is derived by combining momentum conservation between the projectile and target plug with an energy balance that assumes the work required for perforation is independent of impact velocity and equal to its ballistic-limit value. While this formulation is effective for describing the ballistic curve, the evolution of energy absorption is prescribed through the assumed perforation work, therefore, using the same formulation to rationalize energy absorption trends would introduce a circular argument. Later empirical approaches relax these constraints by fitting $a$ and $p$ directly to experimental data [25], but provide limited insight into the physical origin of the observed response and, may yield nonphysical predictions such as residual velocities exceeding the impact velocity when extrapolated beyond the calibration range. A detailed commentary about these two models is provided in Supplementary Note 1. More detailed analytical models have since been developed to capture the $v_i$–$v_r$ response, including energy- and work-based models that partition kinetic-energy loss among deformation and failure mechanisms [26–30], force-balance models that track target and plug motion with prescribed shear or tensile failure criteria [31], and resistance-based models that describe penetration through constitutive resistance and material-strength laws [32, 33] (Refer to Supplementary Note 2 for a broader description of these models). While these formulations provide valuable mechanistic descriptions, they require model-specific quantities associated with energy dissipation, material resistance, or failure criteria. Determining these quantities generally requires knowledge of the material's constitutive and failure response under the relevant impact conditions, and the relevant constitutive and failure properties are often not independently known at the extreme strain rates accessed in LIPIT, reaching $10^6$–$10^9$ $s^{-1}$ [34].

Here, we return to the fundamental quantities of energy and momentum transferred from the projectile during impact. These quantities are determined directly from the measured incident and residual projectile velocities, without prescribing how energy is dissipated, how momentum is partitioned within the target, or how the target fails. We then introduce ballistic-limit-based normalizations to compare projectile–target systems with different masses and geometries. We establish the momentum-transfer trends using high-throughput microprojectile impact experiments, then test their broader applicability using a limited set of in-house macroscale impact experiments together with an expanded compilation of ballistic data from the literature spanning different projectile and target materials, geometries, length scales, and oblique impact conditions. These results show an upper bound on the momentum transfer and specific energy absorption, which we additionally use to provide a physical interpretation of the ballistic limit as the momentum capacity of the target against the corresponding projectile. We examine the mechanisms underlying the observed momentum-transfer trends and subsequently apply the momentum transfer framework to elucidate scale-dependent ballistic-performance comparisons. Together, these results provide a unified basis for interpreting and comparing perforation behavior across impact systems.

## 2. Projectile impact experiments

### 2.1. Experimental design and target preparation

Polystyrene (PS) was selected as a model target material to examine projectile impact while systematically varying the projectile–target geometry, with the silica projectiles treated as near-rigid relative to the PS targets based on $E_{\mathrm{Si}} \approx 72$ GPa and $E_{\mathrm{PS}} \approx 3.2$ GPa. For the microscale impact tests, PS films with thicknesses $h = 0.33$–$6.9$ $\mu$m were combined with three mean projectile diameters, $D = 3.2,\ 8.5$, and 22.4 $\mu$m (see Supplementary Note 6), to produce projectile-to-target thickness ratios of approximately $D/h \approx 3,\ 6$, and 10 (see Fig. 2a). Relative to the smallest microprojectile, the corresponding projectile-size scale factors were $\lambda = D/D_{\mathrm{min}} = 1,\ 2.7$, and 7. A limited set of macroscale gas-gun experiments was additionally performed using 2 mm-diameter borosilicate projectiles and approximately 0.2 mm-thick PS targets, corresponding to $D/h \approx 10$ and a projectile-size scale factor of $\lambda = 625$. Together, the micro- and macroscale experiments provide comparisons across substantially different length scales while retaining comparable projectile–target geometries. The projectile densities used in the analysis, as specified by the manufacturers, are summarized in Table 1.

**Table 1**
Projectile densities used in the micro- and macroscale impact experiments.

| $D$ | $\rho_p$ (kg m$^{-3}$) |
|---|---|
| 3.2 $\mu$m | 2000 |
| 8.5 $\mu$m | 2000 |
| 22.4 $\mu$m | 2560 |
| 2 mm | 2230 |

PS films were fabricated in-house using amorphous PS ($M_w$ = 280,000 g/mol, $T_g$ = 106.4 °C) purchased from Sigma-Aldrich. The PS was dissolved in toluene and spin-coated onto 22 × 22 mm borosilicate cover glasses. The PS concentration and spin-coating speed were varied to obtain the required target thicknesses, with the fabrication parameters for each configuration provided in Supplementary Note 4. The solutions were allowed to dissolve for 12 h before spin coating for 60 s, followed by drying for 6 h.

The film edges were released with a razor blade, and the substrates were immersed in deionized water to detach the PS films from the glass. The freestanding films were collected using a copper loop and transferred onto nickel TEM grids for the microprojectile tests, with diluted adhesive applied around the grid perimeter to secure the films. Film thickness was measured using SEM in combination with focused ion beam sectioning. All film thicknesses were substantially greater than the estimated PS-chain length of approximately 35 nm (Supplementary Note 3), minimizing potential geometric-confinement-induced changes in material behavior [12]. The PS density used throughout the analysis was $\rho_t = 1047$ kg m$^{-3}$, based on the manufacturer-specified density of the starting material.

## 2.2. Microscale projectile impact testing

Figure 1(b) illustrates the LIPIT setup, where the launch pad and PS target were aligned above an inverted microscope (Axio Vert.A1, Zeiss) to visualize the selected microparticle and the target window prior to impact. The launch pads (Fig. 1(b)(i)–(iii)) consisted of borosilicate cover glasses sputter-coated with a 60 nm chromium ablation layer (ACE600, Leica) and a 30 $\mu$m polydimethylsiloxane (PDMS) layer. PDMS (Sylgard 184, Dow Inc.; 10:1 base-to-curing-agent ratio) was spin-coated onto the chromium layer and cured at 200 °C for 1 h. The silica microparticles were suspended in ethanol, vortex-mixed (LP Vortex Mixer, Thermo Scientific), and centrifuged (BenchMate C6V, Oxford) for 60 s; this process was repeated twice to remove surface debris. The cleaned microparticles were then drop-cast onto the cured PDMS and allowed to dry at room temperature.

Individual microparticles were launched using a pulsed Nd:YAG laser (Quanta-Ray, Spectra-Physics; wavelength 1064 nm, pulse width 5–8 ns, pulse energy 0.4 J). Laser ablation of the chromium layer generated rapidly expanding plasma that drove the overlying PDMS layer and accelerated a selected microparticle toward the freestanding PS target (Fig. 1(b)). Projectile velocity was controlled by varying the incident laser power, providing accessible velocities of approximately 100–1000 m/s.

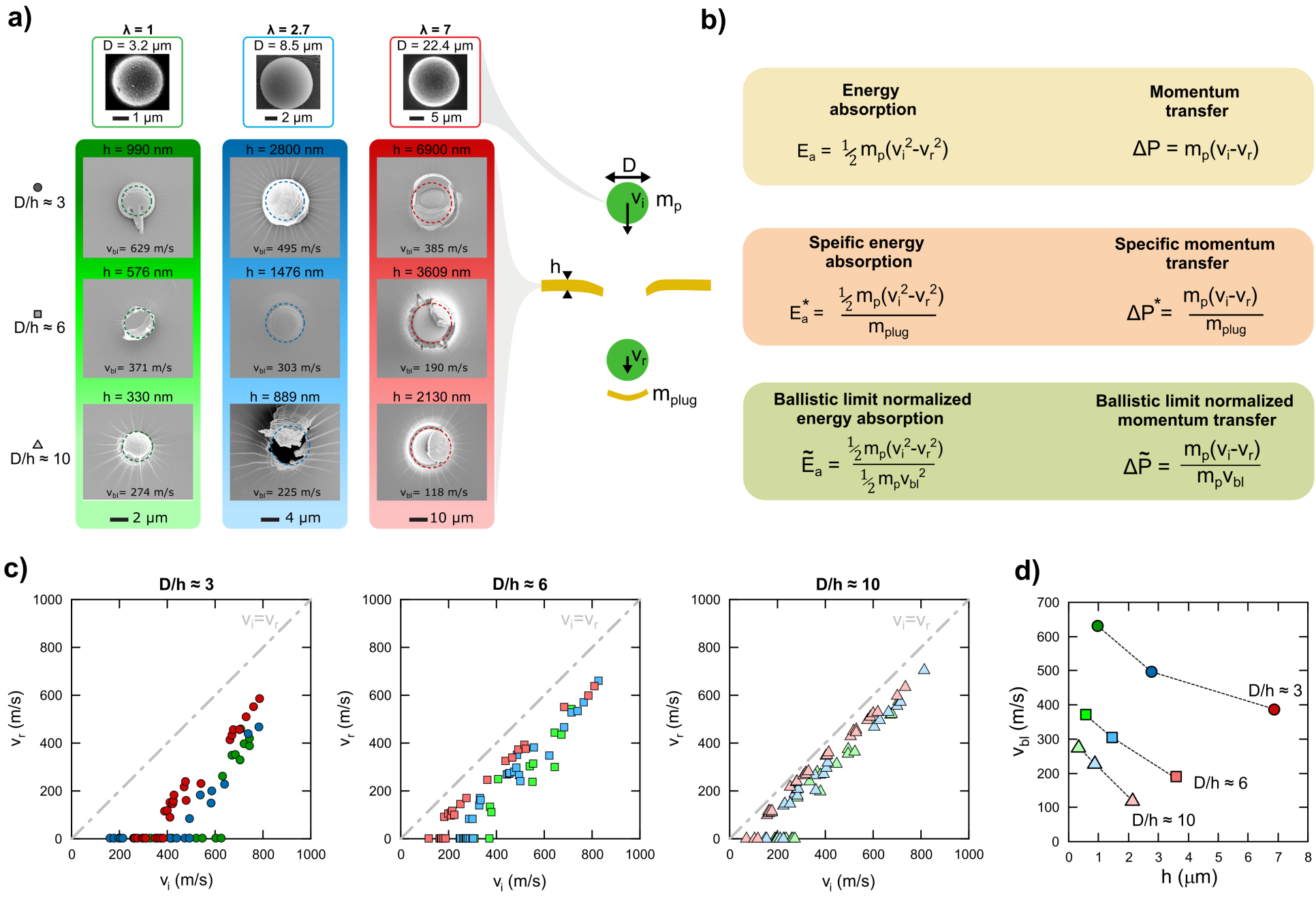

**Figure 2:** Impact experiments at different scales. **(a)** Experimental matrix of nine microscale impact configurations showing the projectile diameter $D$ and target thickness $h$, together with post-impact SEM images illustrating the evolution of deformation morphology from brittle-like fracture at lower ballistic-limit velocities to ductile-like flow at higher ballistic-limit velocities. Each column represents a projectile length scale $\lambda = D/D_{\min}$, and each row represents a $D/h$ ratio. The green–blue–red colors and circle–square–triangle markers uniquely identify each geometry throughout the paper. **(b)** Energy and momentum lost by the projectile are determined from the measured incident and residual velocities and normalized for each configuration, as illustrated. **(c)** Measured residual velocity $v_r$ as a function of incident velocity $v_i$ for $D/h \approx 3,\ 6,$ and 10. The dashed diagonal represents $v_r = v_i$, corresponding to zero projectile velocity reduction during impact. **(d)** Ballistic-limit velocity as a function of target thickness, showing higher $v_{bl}$ at smaller $\lambda$ for configurations with comparable $D/h$. Increasing $h$ at fixed $D$ also increases $v_{bl}$.

Projectile trajectories immediately before and after impact were recorded using a long-working-distance microscope lens (Optem Fusion 12.5:1, Qioptiq) coupled to a monochrome camera (Mako G-234B, Allied Vision). Time-resolved illumination was provided by a pulsed white-light source (SuperK EXTREME 20, NKT Photonics; 350–800 nm) gated by an acousto-optic modulator (Isomet 1250C-848). Light pulses separated by 128.3–257.1 ns generated the multi-exposure side-view images shown in Fig. 1(b)(iv), from which the incident and residual projectile velocities, $v_i$ and $v_r$, were determined directly. The measured trajectories were corrected for aerodynamic deceleration to obtain the projectile velocities immediately before and after impact; the correction procedure is provided in Supplementary Note 5. The PS impact experiments analyzed in this study span incident velocities of approximately 100–800 m/s.

Following impact, the perforated PS targets were examined using SEM to characterize the deformation and fracture morphologies shown in Fig. 2(a). To minimize surface charging, the films were sputter-coated with three successive 3 nm gold layers deposited at 0° and ±10°. Imaging was performed using Zeiss Gemini 300 and Gemini 450 microscopes at an accelerating voltage of 3 kV.

### 2.3. Macroscale projectile impact testing

Figure 1(c) illustrates the macroscale gas-gun setup, where the PS target is mounted on a metal O-ring and the 2 mm projectile is accelerated toward the target using compressed air through a 2.2 mm-inner-diameter launch tube. The launch tube is connected to the compressed-air source through a solenoid valve, and the incident velocity is controlled by the applied pressure. The PS targets were fabricated using the same spin-coating procedure as the microscale films.

Projectile trajectories before and after impact were recorded using a high-speed camera (FASTCAM SA-Z, Photron) at 100,000 frames per second. Incident and residual velocities were extracted using an in-house MATLAB point-tracking routine, with a reference image acquired at the impact plane used for spatial calibration. Impact velocities of approximately 10–200 m/s were accessible with the gas-gun setup.

## 3. Normalized energy and momentum transfer

The measured $v_i$–$v_r$ response across the different microprojectile–target configurations is shown in Fig. 2(c). The $v_r = v_i$ diagonal represents the limiting case of zero projectile deceleration; therefore, increasing separation below this line corresponds to a larger velocity reduction imparted by the target. Across the experimental matrix, lower $D/h$ configurations generally lie farther below this diagonal, indicating stronger target retardation as the target becomes thicker relative to the projectile diameter. Among configurations with comparable $D/h$, smaller $\lambda$ consistently exhibits larger velocity reductions. The measured $v_i$–$v_r$ relation therefore retains a $\lambda$-dependent contribution beyond the explicit geometric ratio $D/h$ in the specific energy absorption formulation.

The same dependence is reflected in the corresponding ballistic-limit velocities shown in Fig. 2(d). For a fixed projectile diameter, increasing the target thickness increases $v_{bl}$; at comparable $D/h$, decreasing $\lambda$ shifts $v_{bl}$ to higher values. The combined variation of $D/h$ and $\lambda$ therefore produces a systematic progression in perforation resistance across the experimental matrix. This progression is accompanied by a pronounced evolution in the failure morphology at the ballistic limit shown in Fig. 2(a). Configurations with larger $\lambda$ and higher $D/h$, located toward the bottom-right of the matrix, predominantly exhibit brittle-like failure characterized by sharp perforation boundaries and radial or tangential crazing. Moving diagonally toward smaller $\lambda$ and lower $D/h$, $v_{bl}$ increases and the deformation becomes increasingly ductile- or flow-like, with greater material stretching and localized deformation around the perforation region. Notably, all of these responses arise from the same PS material system, yet varying $D/h$ and $\lambda$ produces markedly different ballistic-limit velocities and failure morphologies. This transition from brittle- to ductile-like failure suggests that the material response evolves with both projectile–target geometry and absolute scale, likely through the combined effects of rate-dependent polymer behavior, deformation localization, and impact-induced thermal softening.

We examine how this evolution manifests into the energy absorption, $E_a$, and momentum transfer, $\Delta P$ from the projectile. Although the impact velocities are comparable across the experiments, the seven-fold variation in projectile diameter produces substantial differences in projectile mass, $m_p \propto D^3$, causing the dimensional energy and momentum transfers to span several orders of magnitude, as illustrated in Fig. 3. Hence, we normalize the energy absorption and momentum transfer by their corresponding ballistic-limit values, defining $\tilde{E}_a$ and $\Delta\tilde{P}$, respectively, as illustrated in Fig. 2(b). The incident velocity is similarly normalized as $\tilde{v}_i = v_i/v_{bl}$, such that the arrested ($\tilde{v}_i < 1$) and perforated ($\tilde{v}_i > 1$) impacts are cleanly separated with ballistic limit occurring at $\tilde{v}_i = 1$. For arrested impacts, the projectile has no residual velocity ($v_r = 0$); therefore, the kinetic energy and momentum lost by the projectile are simply its incident kinetic energy, $\frac{1}{2}m_p v_i^2$, and incident momentum, $m_p v_i$, respectively. Normalizing these arrested quantities by their corresponding ballistic-limit values removes the configuration-specific mass term. Consequently, the arrested responses for all projectile–target configurations collapse onto a common linear trajectory in the momentum landscape, $\Delta\tilde{P} = \tilde{v}_i$, and a common quadratic trajectory in the energy landscape, $\tilde{E}_a = \tilde{v}_i^2$, as shown in Fig. 3. These arrested trajectories are fully prescribed up to the ballistic limit, beyond which perforation occurs and the resulting momentum and energy transfer are no longer known in advance, making the post-perforation response the regime of interest.

With all impact configurations referenced to their respective ballistic-limit states, the post-perforation evolution of momentum and energy transfer can directly be compared in Fig. 4(b,d). Immediately beyond the ballistic limit, where $\tilde{v}_i \approx 1 + \delta$ and $\delta \to 0^+$, the normalized momentum transfer exhibits a sharp decrease. This reduction is approximately 60% for the $D/h \approx 10$ configurations, whereas a smaller drop of approximately 30% is observed for $D/h \approx 3$. The normalized energy absorption also exhibits a small decrease at the ballistic-limit transition, although the change is less pronounced than in the momentum representation.

As the impact velocity increases further into the perforation regime, the responses diverge considerably with projectile–target geometry. The $D/h \approx 10$ configurations exhibit increasing momentum and energy transfer with $\tilde{v}_i$, whereas the $D/h \approx 3$ configurations show a progressive decrease in momentum transfer accompanied by saturation of the absorbed energy. These contrasting momentum-transfer trends coincide with distinct deformation and failure morphologies observed in the SEM images of the impacted targets. The decreasing-momentum and energy-saturating cases exhibit extensive flow-like deformation and thermally softened features around the perforation region (see Fig. 4(c)), whereas configurations exhibiting increasing momentum and energy transfer retain predominantly brittle-like features, including radial and tangential crazing and sharply fractured perforation boundaries (see Fig. 4(a)). Thus, even within the same PS material system, substantially different momentum- and energy-transfer trajectories emerge as the dominant deformation and failure response evolves.

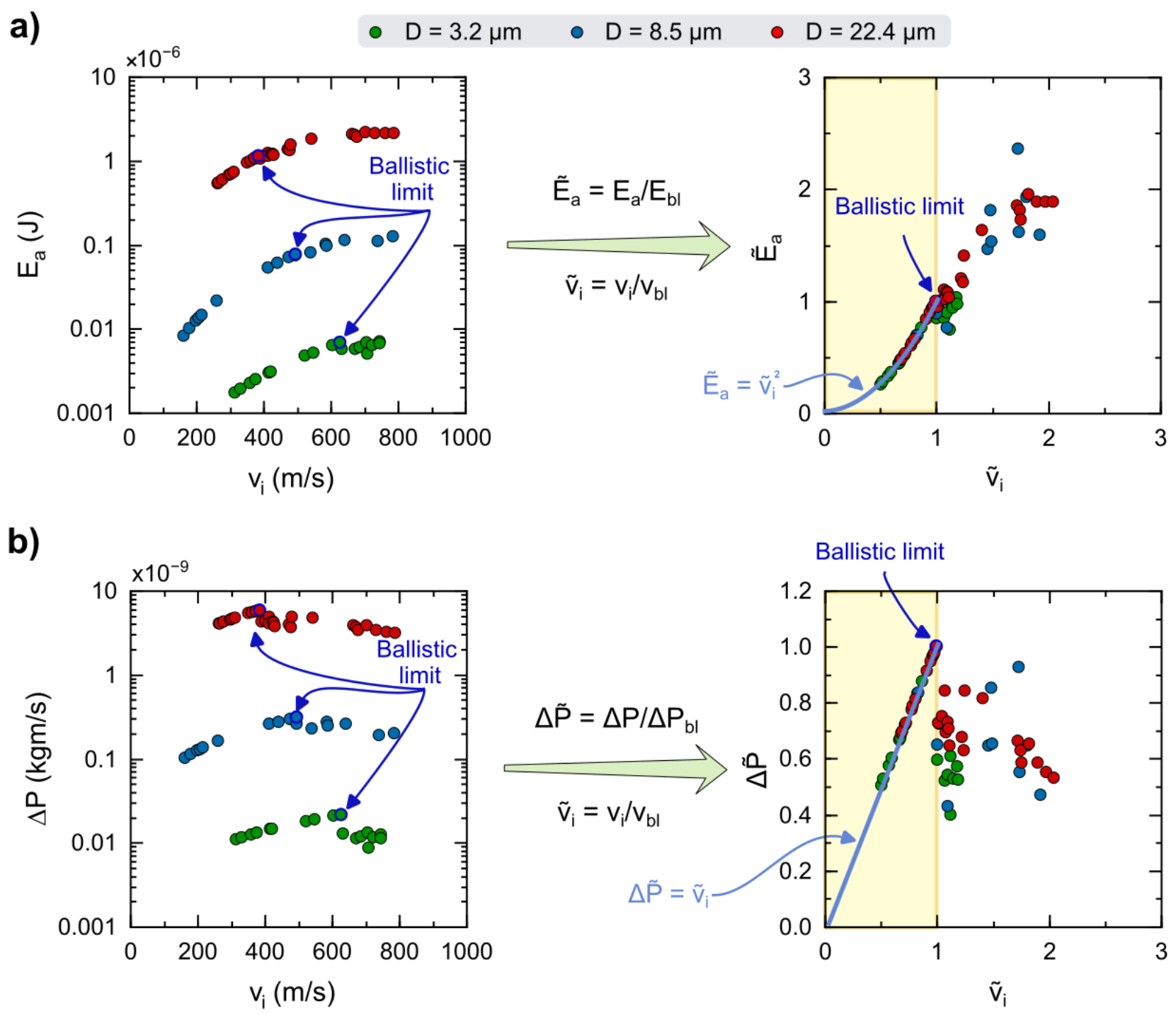

**Figure 3: Ballistic-limit normalization of energy and momentum transfer.** (a) Energy absorption before and after normalization by $E_{bl}$. (b) Momentum transfer before and after normalization by $P_{bl}$. Arrested impacts collapse onto common trajectories until reaching the ballistic limit at $(1, 1)$.

Despite these differences, a striking common feature appears in the momentum landscape: all perforated cases remain below the momentum transferred at the ballistic limit, such that

$$\Delta \tilde{P} < 1. \tag{3}$$

This relation maps to the energy absorption domain as

$$\tilde{E}_a < 2\tilde{v}_i - 1, \tag{4}$$

which appears as the linearly increasing upper envelope in Fig. 4(d). These experimental observations highlight that the same impact response is represented by a fixed upper ceiling in momentum space but an expanding envelope in the energy landscape. Interestingly, a similar maximum in momentum transfer at the ballistic limit was reported for macroscale steel-plate perforation [35], but this observation received limited subsequent attention. Two materially distinct systems subjected to impact at vastly different length and strain-rate scales exhibit the same momentum-transfer behavior, prompting us to question how broadly this response extends across other projectile–target systems.

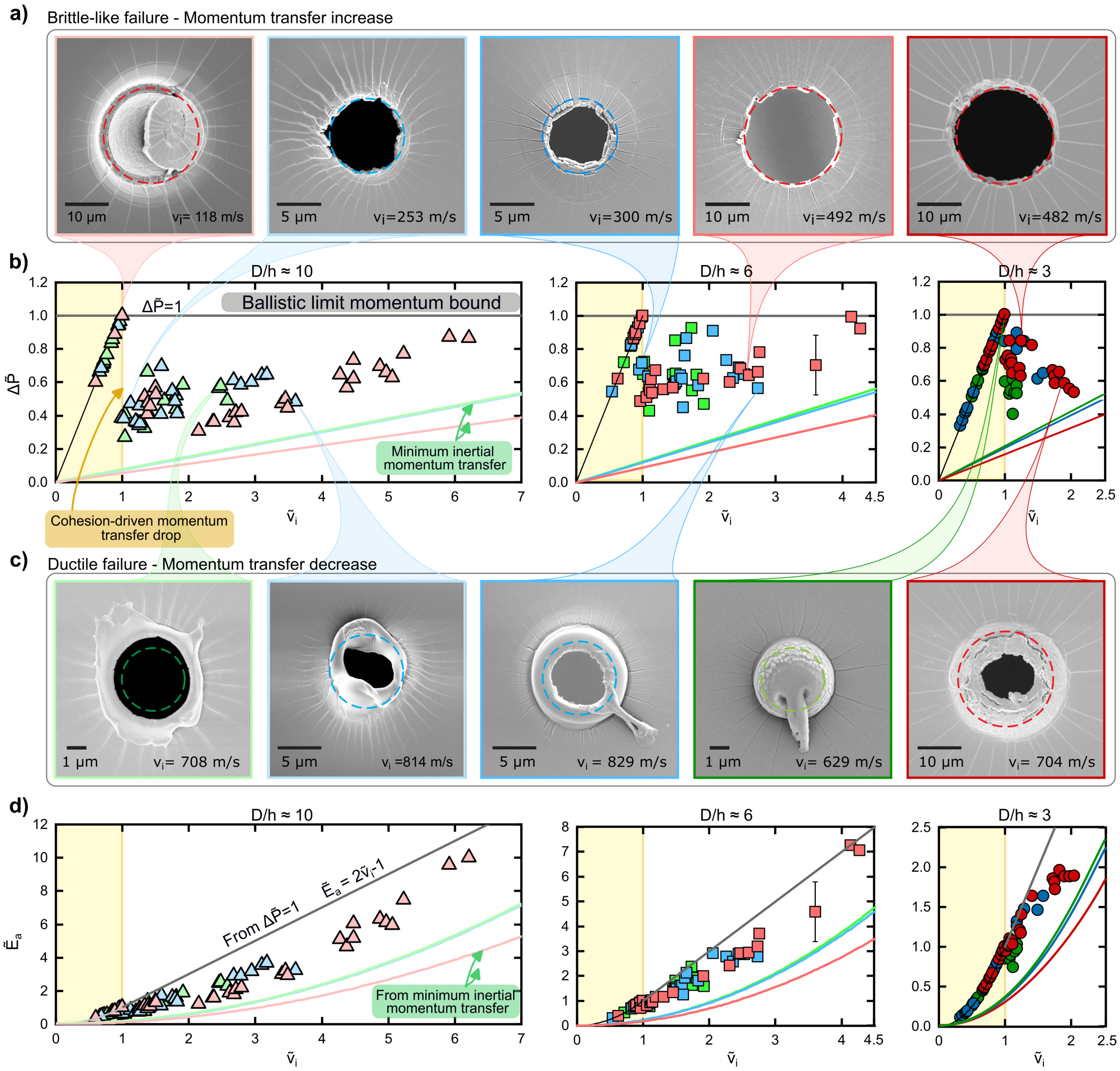

**Figure 4:** Normalized momentum and energy transfer trends and the corresponding failure mechanisms. **(a)** SEM images of brittle-like failures showing radial/tangential crazes, sharp boundaries, and minimal thermal softening, typically occurring at lower $\widetilde{v}_i$ and higher $D/h$ ratios where momentum transfer increases. **(b)** Normalized momentum versus normalized impact velocity for $D/h \approx 10$, 6, and 3. Yellow regions indicate arrested cases ($v_i < v_{bl}$), where all data collapse to $\Delta\widetilde{P} = \widetilde{v}_i$. All perforated cases lie below the ballistic-limit momentum. The minimum inertial momentum transfer for each geometry appears as colored solid lines. **(c)** SEM images of ductile failures with thermally softened features, occurring at higher $\widetilde{v}_i$ and lower $D/h$ ratios where momentum transfer decreases. **(d)** Corresponding normalized energy-absorption plots, showing the momentum-transfer bound that simplifies to $\widetilde{E}_a = 2\widetilde{v}_i - 1$ and the inertial lower bound. Representative error bars for measurement uncertainties are shown for $D/h \approx 6$ in (b) and (d) (see Supplementary Note 6).

## 4. Unifying bounds and re-imagining the ballistic limit using momentum transfer

A broad survey of 16 independent studies, comprising more than 640 data points across metals, polymers, composites, sandwich panels, and reinforced concrete slabs, reveals the same behavior: the momentum transferred to the target peaks at the ballistic limit and never exceeds the ballistic-limit value [7, 12, 25, 29, 32, 36–46]. This universality persists over thicknesses ranging from nanometers to hundreds of millimeters, projectile diameters from microns to centimeters, and impact geometries spanning spherical, blunt, ogive, and conical projectiles. Notably, the same scaling behavior is observed under both normal and oblique impact conditions (Fig. 5a). We did not encounter a single perforation dataset in which the momentum transfer exceeded the ballistic-limit value, suggesting that this constraint is deeply embedded in the mechanics of projectile–target interactions rather than specific to a particular material or configuration.

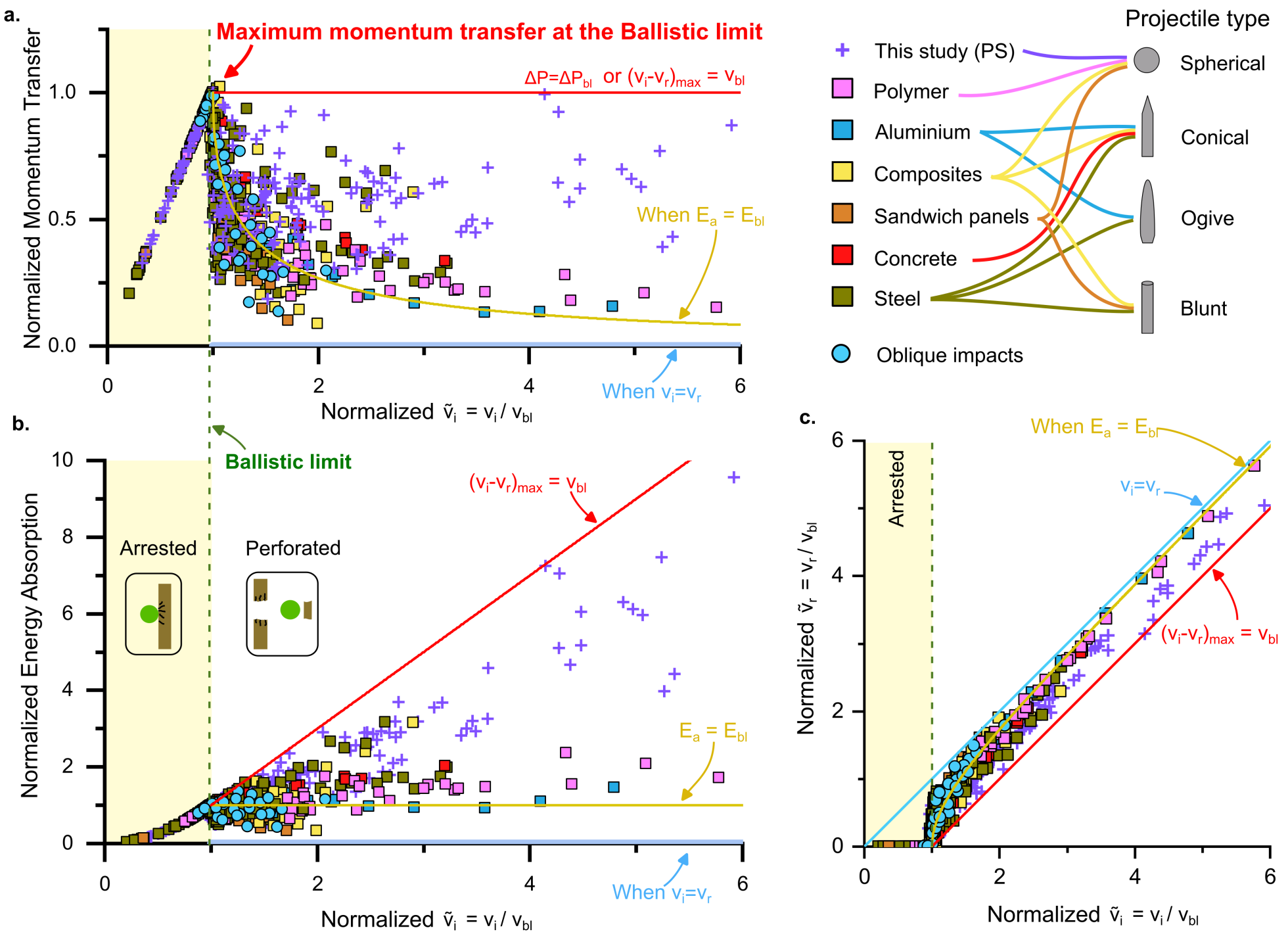

**Figure 5:** Unified impact-response trends revealed through momentum-based normalizations. **(a)** Normalized momentum transfer, **(b)** normalized energy absorption, and **(c)** normalized residual velocity as functions of normalized incident velocity. Across all datasets, the maximum momentum transfer consistently occurs at the ballistic limit, while the energy absorbed during penetration may increase or decrease relative to the ballistic-limit value. The figure includes 234 impact cases from the present study and 642 cases compiled from the literature. Each color denotes a different target material, and the corresponding projectiles are indicated. Circle markers represent oblique impacts. The yellow line represents the case $E_a = E_{bl}$ in all three landscapes.

This universality in the experimental impact data marks an important distinction in how projectile impact mechanics is currently understood. Energy and momentum transfer arise from the same projectile kinematics, yet only the momentum response exhibits a consistent extremum at the ballistic limit. The energy-absorption landscape in Fig. 5(b) contains numerous perforating impacts for which $E_a < E_{bl}$, which are difficult to reconcile with interpretations that define perforation through a required energy or work threshold discussed in Supplementary Note 1. Thus, we offer a physically grounded definition of the ballistic limit: *the ballistic limit corresponds to the maximum momentum that a given target can sustain without perforation—the target's momentum capacity—from a specific projectile collision.* The momentum-capacity definition provides a quantifiable, mechanics-based description of the ballistic limit, rather than current practice of identifying it solely as a phenomenological test metric from the experimentally observed transition between arrest and perforation or through statistical measures such as the $V_{50}$ limit, which corresponds to a 50% probability of the target stopping the projectile.

Figure 5(c) further illustrates how all impact data are bounded within a finite region. The line $v_i = v_r$ defines the zero-momentum-transfer edge, whereas the momentum bound defines the opposite edge, $v_i - v_r < v_{bl}$. An immediate consequence of the momentum bound is that *no perforating target can reduce the impacting projectile velocity by more than its ballistic-limit velocity*. The ballistic-limit velocity therefore has direct mechanical significance for protective-system design beyond marking the arrest-to-perforation boundary, because increasing $v_{bl}$ raises the maximum velocity reduction that a target may impart to a projectile even when perforation occurs.

## 5. Rationalizing the momentum-transfer trends

Having established experimentally that the ballistic limit is consistently marked by a maximum in momentum transfer, and that the observed momentum-transfer trends are closely linked to the underlying failure mechanisms, we next examine the underlying physical mechanisms that give rise to these distinct responses. The evolution of the projectile's momentum during collision can be represented by the impulse,

$$\Delta P = \int_0^t F\, dt = m_p(v_i - v_r), \tag{5}$$

where $F$ is the magnitude of the retardation force acting on the projectile during the short contact duration associated with perforation. The magnitude of this impulse is governed by the target's immediate region of influence, which actively responds to the impact by generating resistive forces through its deformation mechanisms. For high-speed collisions, this region of influence is determined by the wave speed in the target material and the dominant deformation and failure mechanisms. The impulse generated by the target in response to the striking projectile can arise from several interacting mechanisms. Here, we focus on two influential contributions: target inertia and material cohesion. The former represents the impulse required for the instantaneous acceleration of the mass within the region of influence, while the latter represents the internal stresses generated when deforming the same region. For simplicity, we decouple these two contributions by considering the impact response of a cohesionless mass responding only through inertia and an idealized zero-density material responding purely through its constitutive law.

In a cohesionless target (e.g., a granular system), the target readily separates and disperses upon impact due to the absence of interparticle cohesive forces, highlighting the importance of material cohesion in resisting and ultimately arresting a colliding projectile. As impact velocity increases, the momentum delivered by the projectile induces higher internal stresses within the target's region of influence, and this response is highly nonlinear, governed by elasto-plastic wave propagation, failure initiation and evolution, and adiabatic heating from deformation and friction at the projectile–target interface.

The initial stress wave rapidly disperses through the target, causing each material point in the region of influence to undergo a loading–unloading cycle with a corresponding peak stress state. To simplify the complex failure processes involved, we introduce a conceptual critical material point whose state determines whether perforation occurs or not. Just before the ballistic limit, this point approaches but does not exceed its failure threshold, allowing the target to unload and produce an extended impulse curve shown by the pink curves in Fig. 6(a). A marginal increase in impact velocity causes the critical material point to exceed this threshold (blue in Fig. 6(a)), resulting in plug detachment and a sharp reduction in momentum transfer due to the loss of the unloading phase. If the critical point fails later in the impact sequence, the magnitude of momentum loss is reduced—explaining the experimental trends observed in Fig. 4(b). Lower $D/h$ targets tend to undergo thermal softening and polymer chain elongation, delaying failure, whereas higher $D/h$ targets exhibit brittle-like failure, where critical failure can occur much earlier, truncating a larger portion of the impulse and leading to a more pronounced drop in transferred momentum. The difference in the momentum-transfer drop at the ballistic limit therefore serves as an indicator of a shift in the target's dominant failure mode.

As the impact velocity increases beyond the ballistic limit, the material deforms at higher strain rates, which can increase the resistive forces generated by the rate-dependent response of the polymer. Simultaneously, the impact duration may shorten, leading to a more localized region of influence (see Supplementary Note 9 for estimates). Although these competing effects can either amplify or suppress cohesion-driven momentum transfer, experimental data consistently show that all post-perforation momentum transfers remain below the value observed at the ballistic limit ($\Delta\tilde{P} < 1$). However, for targets with high diameter-to-thickness ratios ($D/h$), momentum transfer increases monotonically with impact velocity and, at the highest tested velocities, approaches the ballistic limit—raising the question of whether the observed momentum bound could be violated.

To address this question, we consider an alternative mechanism of momentum transfer: the inertial response in a cohesionless target. Momentarily at impact, the mass within the region of influence is accelerated to match the projectile's velocity profile (Fig. 6(b)), reflecting the instantaneous transfer of momentum. This leads to a minimum inertial momentum transfer,

$$\Delta P_{\min} = m_p(v_i - v_r) = m_{\text{plug}} v_r, \tag{6}$$

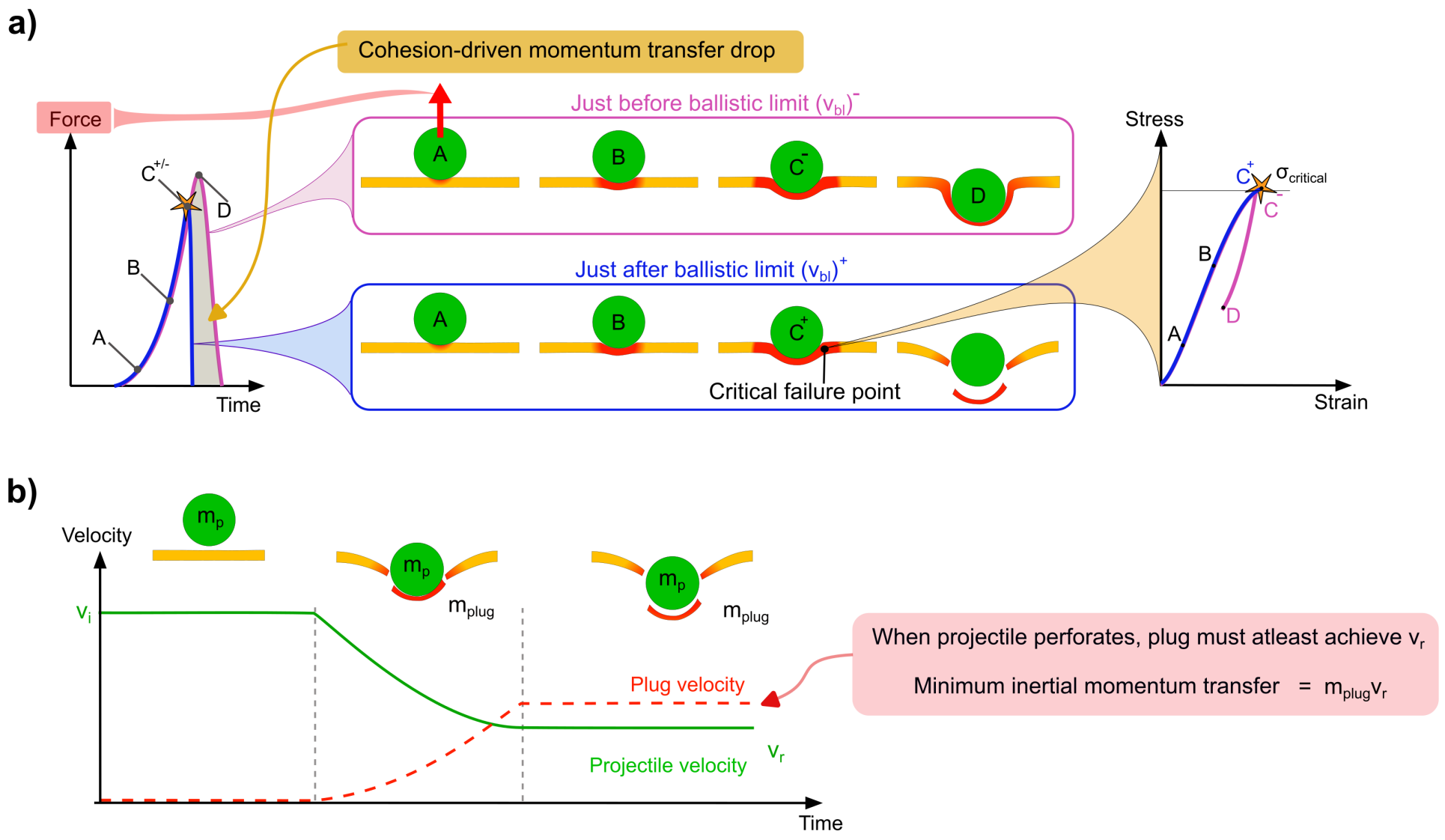

**Figure 6:** **(a)** Cohesion-driven momentum-transfer drop illustrated using impulse curves immediately before and after the ballistic limit. The stress–strain response of a critical failure point illustrates how the pre-limit case survives the peak stress and produces a larger momentum transfer. **(b)** In brittle-like failure, the minimum momentum transfer corresponds to plug ejection at $v_r$.

implying that material points within the ideal plug must be displaced at least at the residual velocity $v_r$ as the projectile exits the target. Resolving the target and projectile momentum changes gives

$$\Delta \tilde{P}_{\min} = \zeta \tilde{v}_i, \qquad \tilde{E}_{a,\min} = \zeta(2-\zeta)\tilde{v}_i^2, \tag{7}$$

where $\zeta = \frac{m_{\text{plug}}}{m_{\text{plug}}+m_p} < 1$.

The corresponding derivation is provided in Supplementary Note 10, and the resulting lower bounds are plotted in Fig. 4(b) and Fig. 4(d). This analysis explains the increasing trend of momentum transfer with $\tilde{v}_i$ and indicates that the inertial contribution alone could, in principle, exceed the momentum bound ($\Delta \tilde{P}_{\min} > 1$) when $\tilde{v}_i > \frac{1}{\zeta}$, although such conditions lie well beyond the ballistic testing regime explored here.

While momentum transfer increases with velocity for larger $D/h$ geometries, lower $D/h \approx 3$ targets show the opposite trend, accompanied by energy-absorption saturation. This behavior coincides with pronounced plastic deformation observed in post-mortem SEM images, suggesting substantial adiabatic heating and thermal softening. At these elevated temperatures, increased chain mobility allows polymer chains to squeeze and slip past one another, enabling the projectile to perforate the target without accelerating the ideal plug mass to $v_r$. This transition in deformation mechanism is marked by the saturation of energy absorption, which results in a decreasing momentum-transfer trend: $\Delta P \sim \frac{E_{\text{sat}}}{v_i}$ (see Supplementary Note 11).

These findings demonstrate that a single constitutive framework can yield fundamentally different ballistic responses depending on the dominant deformation and failure mechanisms. By directly tracking momentum and energy transfer, we uncover mechanistic insights without relying on traditional assumptions such as a closed system, which often underpin continuum or penetration models. While the present discussion focuses on cohesive and inertial momentum-transfer mechanisms, these are not the only mechanisms contributing to the projectile–target interaction. Future investigations should aim to characterize additional contributions—such as interfacial friction—and explore the effects of other energy-saturation phenomena, including pressure-induced liquefaction or plasma formation under extreme impact conditions that may become important for higher impact velocities beyond what is investigated here.

## 6. Revisiting ballistic performance comparisons through momentum transfer

With the momentum bound rationalized, we next examine how it maps into the energy and specific-energy landscapes used to compare ballistic performance. We first establish how the same projectile–target configurations are represented by $E_a$ and $E_a^*$, before using the momentum bound to interpret the resulting trends. For the polystyrene targets studied here, larger projectile diameters lead to substantially higher absorbed energies $E_a$, with values spanning a wide range across scales (see Fig. 7(a)), consistent with the strong projectile-mass dependence $m_p \propto D^3$. At a fixed projectile diameter, increasing target thickness intuitively increases the absorbed energy. Normalizing the energy absorption by the plug mass substantially compresses this wide spread in $E_a$. Furthermore, the normalized representation shows that thinner targets exhibit higher $E_a^*$ and, at comparable $D/h$, configurations with smaller $\lambda$ exhibit higher $E_a^*$ (Fig. 7(c)). At first glance, this suggests that thinner targets at smaller $\lambda$ provide superior ballistic performance, even within the same material system. Note that Fig. 7 includes only perforated cases, since arrested impacts follow a predefined arrested curve; the complete dataset including arrested cases is provided in Supplementary Note 7.

We next examine how the momentum bound is represented within the energy landscape. All arrested impacts populate the trajectory $\hat{E}_a = \frac{1}{2} m_p v_i^2 = E_i$ (a hat denotes an arrested case with $v_r = 0$), and the momentum bound maps onto this landscape as a straight line tangent to the arrested trajectory at the ballistic limit (see Fig. 7(b)). This tangent is described by

$$E_a = \Delta P_{bl} v_i - E_{bl}, \tag{8}$$

where the slope corresponds to the momentum capacity $\Delta P_{bl}$ and the intercept is set by the ballistic-limit energy $E_{bl}$ (see Supplementary Note 8). The tangent is indicated by an arrow, and its starting point, the ballistic limit, is marked by a black circle for each diameter–thickness combination.

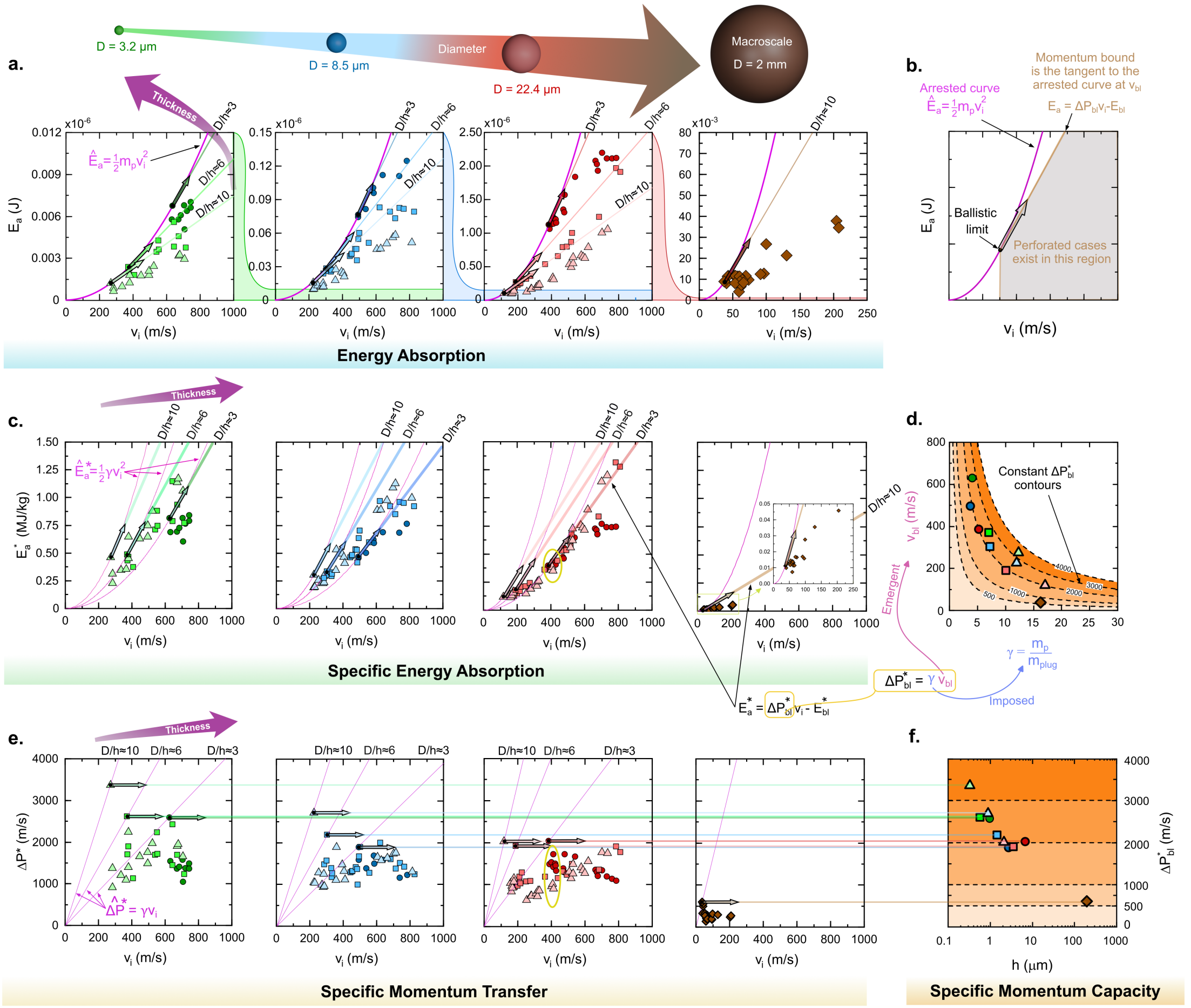

**Figure 7:** Interpreting energy-absorption metrics using the momentum bound. **(a)** Variation of $E_a$ with $v_i$ for $D = 3.2$ $\mu$m, 8.5 $\mu$m, and 22.4 $\mu$m, along with macroscale testing at $D = 2$ mm. The pink curves represent arrested cases, and the arrows represent the momentum bound. **(b)** The momentum bound is tangent to the arrested curve at the ballistic limit. The shaded area represents the region that perforated samples can populate. **(c)** Variation of $E_a^*$ with $v_i$, together with the momentum bounds. **(d)** The gradient of the bound, $\Delta P_{bl}^*$, is the specific momentum capacity that separates the imposed ($\gamma$) and emergent ($v_{bl}$) parameters. **(e)** Variation of $\Delta P^*$ with $v_i$; the horizontal arrows represent the momentum bound, $\Delta P_{bl}^*$. **(f)** Variation of $\Delta P_{bl}^*$ with $h$ for polystyrene targets. Yellow circles in (c) and (e) identify the same impact events, with contrasting ordering based on the metric (see Supplementary Note 12).

The arrested $\hat{E}_a$ trajectories shift upward with increasing projectile diameter according to the scaling $\hat{E}_a \propto D^3 \propto \lambda^3$. Because this trajectory increases monotonically with velocity, $\frac{d\hat{E}_a}{dv_i} = m_p v_i > 0$, a higher ballistic-limit velocity on the same arrested curve necessarily produces a steeper tangent. Together, these relations rationalize the scale variation in absorbed energy with projectile diameter and the steeper $E_a$–$v_i$ trends associated with increasing target thickness at a given diameter through the corresponding increases in ballistic-limit velocity, as shown in Fig. 2(d).

This tangent-based interpretation of the $E_a$ landscape extends directly into the specific energy-absorption domain, where the projectile mass $m_p$ is replaced by the mass ratio $\gamma = m_p/m_{\text{plug}} \propto \frac{\rho_p D}{\rho_t h}$. Consequently, the unique arrested trajectory for a given projectile in the $E_a$ landscape diverges into three separate curves in the $E_a^*$ landscape as the target thickness changes. Because the experimental matrix in Fig. 2(a) was designed using fixed $D/h$ ratios, the corresponding configurations share similar $\gamma$ values and therefore produce approximately equivalent arrested trajectories across the three projectile diameters. Minor deviations arise for the $D = 22.4\,\mu$m case due to the different silica projectile density. The momentum bound similarly transforms to

$$E_a^* = \Delta P_{bl}^* v_i - E_{bl}^*, \tag{9}$$

where the slope is given by the *specific momentum capacity*, $\Delta P_{bl}^* = \gamma v_{bl}$, and the intercept is set by $E_{bl}^* = \frac{1}{2}\gamma v_{bl}^2$.

Extending the same mass normalization into the momentum-transfer landscape produces the specific momentum-transfer relation,

$$\Delta P^* = \gamma(v_i - v_r), \tag{10}$$

plotted in Fig. 7(e). Here, the arrested trajectories remain linear while the momentum bounds appear as horizontal lines for each geometry. For these reference trajectories, the transformation from the specific energy to the specific momentum landscape can be understood through their velocity derivatives: the quadratic arrested trajectories in Fig. 7(c) become linear in Fig. 7(e), while the momentum-bound tangents transform into horizontal lines. Notably, the height of each horizontal bound corresponds to the same specific momentum capacity that defines the slope of the energy bound in the $E_a^*$ landscape.

Comparing perforated-target performance using specific energy absorption and specific momentum transfer reveals a significant inconsistency: the two metrics may lead to contradictory conclusions about ballistic performance. For example, focusing on the data marked by the yellow circles for the $D = 22.4\,\mu$m projectiles (see Fig. 7(c) and (e)), the specific energy representation suggests that all three target thicknesses exhibit similar performance at $v_i = 400\,$m/s. In contrast, the specific momentum representation produces a clear ordering in performance that increases from the thinnest to the thickest target. Similar inconsistencies are observed across the other projectile diameters and velocity ranges, see Supplementary Note 12, demonstrating that the choice of metric alone can alter the apparent ordering of ballistic performance. This raises the fundamental question of which descriptor provides a consistent basis for comparing different target–projectile configurations.

The universal momentum bound identified earlier provides a mechanistic explanation for this contradiction. In both the specific energy and specific momentum landscapes, the momentum bound defines the accessible performance regime that a given target–projectile configuration can attain at a particular velocity. This bound is governed by two quantities: the imposed projectile–target mass ratio, $\gamma$, and the emergent target response through the ballistic-limit velocity, $v_{bl}$. As the imposed condition is varied, the emergent response evolves accordingly, producing the inverse $\gamma$–$v_{bl}$ relationship observed in Fig. 7(d). In the specific momentum landscape, the accessible regime is determined solely by the specific momentum capacity, $\Delta P^*_{bl} = \gamma v_{bl}$, which appears directly as the specific momentum bound for a given target–projectile configuration. As a result, the specific momentum capacity can evolve freely as the imposed condition $\gamma$ is varied, assuming higher, lower, or equivalent values depending on the projectile–target interaction. For example, all three targets corresponding to $D = 22.4\,\mu$m projectiles (red markers) exhibit similar specific momentum capacities of $\Delta P^*_{bl} \approx 2000\,\mathrm{m/s}$, indicating that changes in $\gamma$ produce little change in the performance of these configurations. In contrast, the specific momentum capacity for $D = 3.2\,\mu$m projectiles increases from approximately $2600\,\mathrm{m/s}$ to $3400\,\mathrm{m/s}$ as the target thickness is reduced from 990 nm to 330 nm.

In contrast, the specific energy landscape systematically reorganizes the momentum bound as the imposed condition $\gamma$ is varied. This behavior arises because the momentum bound is represented by a tangent whose slope and intercept are both determined by the same pair of quantities, $\gamma$ and $v_{bl}$. Because both geometric characteristics of the bound evolve simultaneously, changes in $\gamma$ produce a different energy envelope and therefore a different accessible specific energy regime. For the polystyrene data, this manifests as a systematic offset of the energy bounds, with thinner targets—and therefore higher $\gamma$ values—consistently accessing the highest specific energy regimes at a given velocity, as clearly illustrated in Fig. 7(c). This $\gamma$-dependent organization of the momentum bounds explains the contradiction between the specific energy and specific momentum representations: configurations with different imposed conditions are systematically assigned different accessible energy regimes. For this reason, specific momentum transfer provides a more consistent basis for comparing ballistic performance across varying $\gamma$, whereas the specific energy representation systematically favors higher-$\gamma$ configurations. Consistent with this interpretation, cases with similar $\gamma$ values, such as geometrically similar $D/h$ configurations tested with different projectile diameters, produce consistent rankings in both the specific energy and specific momentum representations.

The significance of the specific momentum capacity becomes even clearer when considering multilayered targets. We consider an idealized $n$-layer system in which a projectile decelerates sequentially through identical layers, with each layer transferring the maximum momentum permitted by its ballistic limit. As additional layers are introduced, the effective target mass increases such that $\gamma \propto 1/n$. Accounting for the progressive deceleration of the projectile as it perforates each layer, with the residual velocity carried forward as the incident velocity for the subsequent layer, the effective ballistic-limit response evolves as $v_{bl} \propto n$ (see Supplementary Note 13 and Fig. 8(a)). Consequently, the specific momentum capacity, $\Delta P^*_{bl} = \gamma v_{bl}$, remains unchanged across the multilayered system, as shown by the constant gradients of the momentum-bound tangents in the specific energy landscape in Fig. 8(b). Furthermore, this analysis again demonstrates how the specific energy metric elevates the maximum accessible specific energy by $E^*_{bl} = \frac{1}{2}\gamma_{(n=1)} v^2_{bl}$ for each additional layer because of the embedded geometric scaling effects rather than a change in the material response.

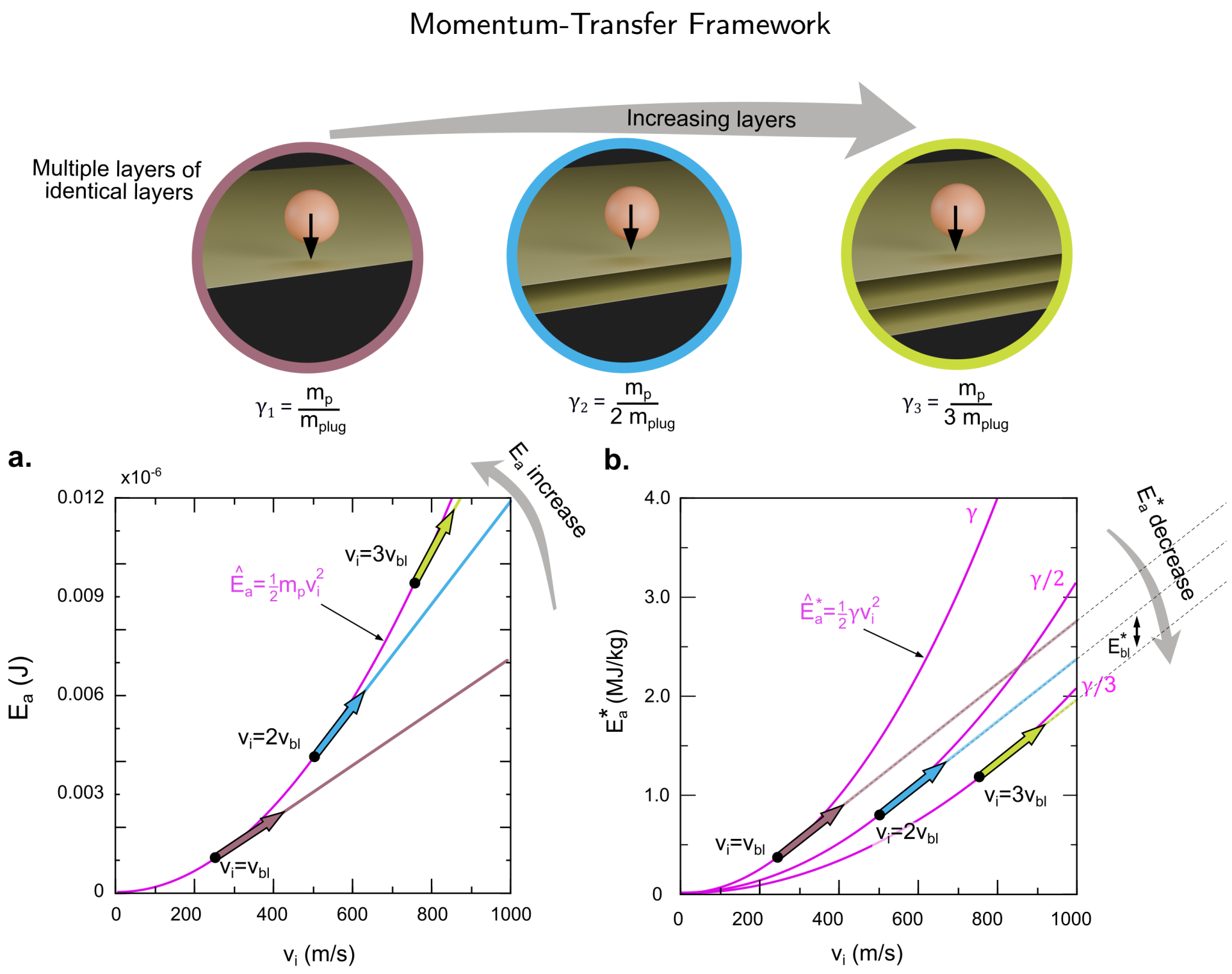

**Figure 8:** Evolution of the ballistic performance of multiple identical layers. **(a)** $E_a$ and **(b)** $E_a^*$ comparisons for a multilayered target impacted by the same projectile, where each layer has the same material performance (see Supplementary Note 13). The $E_a$ bound increases with the number of layers, while the $E_a^*$ bound shifts downward by $E_{bl}^*$ while maintaining the same gradient.

Beyond providing a more consistent basis for comparing perforation responses, the specific momentum capacity, $\Delta P_{bl}^*$, offers a velocity-independent measure of ballistic performance. Because this quantity corresponds to the maximum attainable specific momentum transfer, it provides a compact description of the overall momentum-transfer capability of a target–projectile configuration. Fig. 7(f) visualizes this quantity through contours of constant specific momentum capacity, providing a framework to examine how ballistic response evolves as target thickness and projectile size are varied. For the polystyrene data, most microprojectile configurations organize around a common specific momentum capacity of approximately 2000 m/s. However, targets thinner than 1 $\mu$m consistently occupy higher contours, reaching specific momentum capacities approaching 3400 m/s. Unlike the systematic offsets observed in the specific energy landscape, this enhancement reflects a change in the ballistic response captured by the specific momentum capacity and may be associated with nanoscale confinement, strain-rate effects, or other thickness-dependent deformation mechanisms. At the opposite extreme, the macroscale target occupies a substantially lower contour, with a specific momentum capacity near 500 m/s, indicating a pronounced shift in the underlying deformation and failure response.

More broadly, this work shifts the focus from comparing ballistic performance through specific energy absorption to understanding it through specific momentum transfer. In doing so, it reframes the central objective of ballistic materials research. Rather than seeking materials that simply absorb more energy, the challenge becomes understanding what governs momentum-transfer capacity and how that capacity can be most effectively realized at perforation velocities. Answering these questions will require connecting material constitutive behavior, deformation, and failure mechanisms directly to momentum-transfer pathways. Such a framework could ultimately enable predictive design methodologies and material-specific scaling laws that span length scales, target geometries, and impact conditions.

## 7. Summary

We show that, across the projectile–target systems examined, momentum transfer consistently exhibits its maximum at the ballistic limit, establishing a universal momentum bound for perforating impacts. This observation provides a physically grounded definition of the ballistic limit as the maximum momentum that a given target can sustain without perforation—the target's momentum capacity—for a specific projectile collision. The same bound also places a direct limit on the projectile velocity reduction achievable by a target during a perforating impact such that the projectile velocity cannot be reduced by more than the target's ballistic-limit velocity. These findings give the ballistic limit direct mechanical significance beyond its conventional identification as the transition between projectile arrest and perforation.

The momentum bound also reveals an important limitation in ballistic-performance comparisons. Specific energy absorption and specific momentum transfer can produce contradictory rankings because the specific-energy landscape remains strongly influenced by the imposed projectile–target mass ratio, which can shift the accessible energy regime and inflate the apparent performance of certain configurations. Specific momentum capacity provides a more direct measure of how the ballistic response itself evolves as projectile size, target thickness, and scale are varied. These insights also provide a new basis for analytical modeling. Among the perforation models examined, the ballistic limit is generally introduced through energy, work-to-perforation requirements, or stress- and strain-based failure criteria; none uses the experimentally observed maximum in momentum transfer as the ballistic criterion. Tracking how momentum capacity evolves with material response, geometry, and scale therefore offers a path toward analytical models that can more directly rationalize scale-dependent ballistic response.

The physical mechanisms controlling momentum transfer during perforation are not yet understood in their entirety. The present results show how cohesion, failure, inertia, thermal softening, and energy saturation can generate markedly different momentum-transfer trends, while additional mechanisms such as interfacial friction may also become important. Developing a more complete understanding of these momentum-transfer pathways should enable protective systems to be designed to maximize momentum capacity and operate as close as possible to the momentum bound before perforation. The momentum bound therefore provides both a unifying description of projectile impact and a physical design target for developing lighter and more effective protective materials.

## A. Supplementary Data

Supplementary material related to this article is attached separate.

## Data availability

All the data of this study are provided in main text and figures of the manuscript and the supplementary information. Additional information and other findings of this study are available from the corresponding author upon request.

## Acknowledgements

We acknowledge the financial support from the U. S. Office of Naval Research under PANTHER award number N00014-24-1-2200 through Dr. Timothy Bentley, the Wisconsin Alumni Research Foundation's Accelerator Grant, and the Vilas Faculty Early Career Investigator Award from the Office of the Provost of the University of Wisconsin-Madison.

## Author contributions

YD, NJ, and RT developed the concepts. NJ and JC designed and built the experiments and fabricated the materials. NJ, JC, and AK performed the experiments. YD, NJ, and AK analyzed the data. YD and NJ developed analytical models. YD, NJ, and RT interpreted the results and wrote the manuscript with input from all authors. RT conceived and supervised the entire project.

## Declaration of competing interest

Authors declare no competing interests.

# Supplementary materials for "Momentum-Transfer Framework Unifies High-Velocity Impact and Failure Across Materials, Geometries, and Scales"

Yasara Dharmadasa, Nicholas Jaegersberg, Ara Kim, Jizhe Cai, Ramathasan Thevamaran

**Table of contents**

## Supplementary Note 1: A short summary of Recht-Ipson impact modeling methods

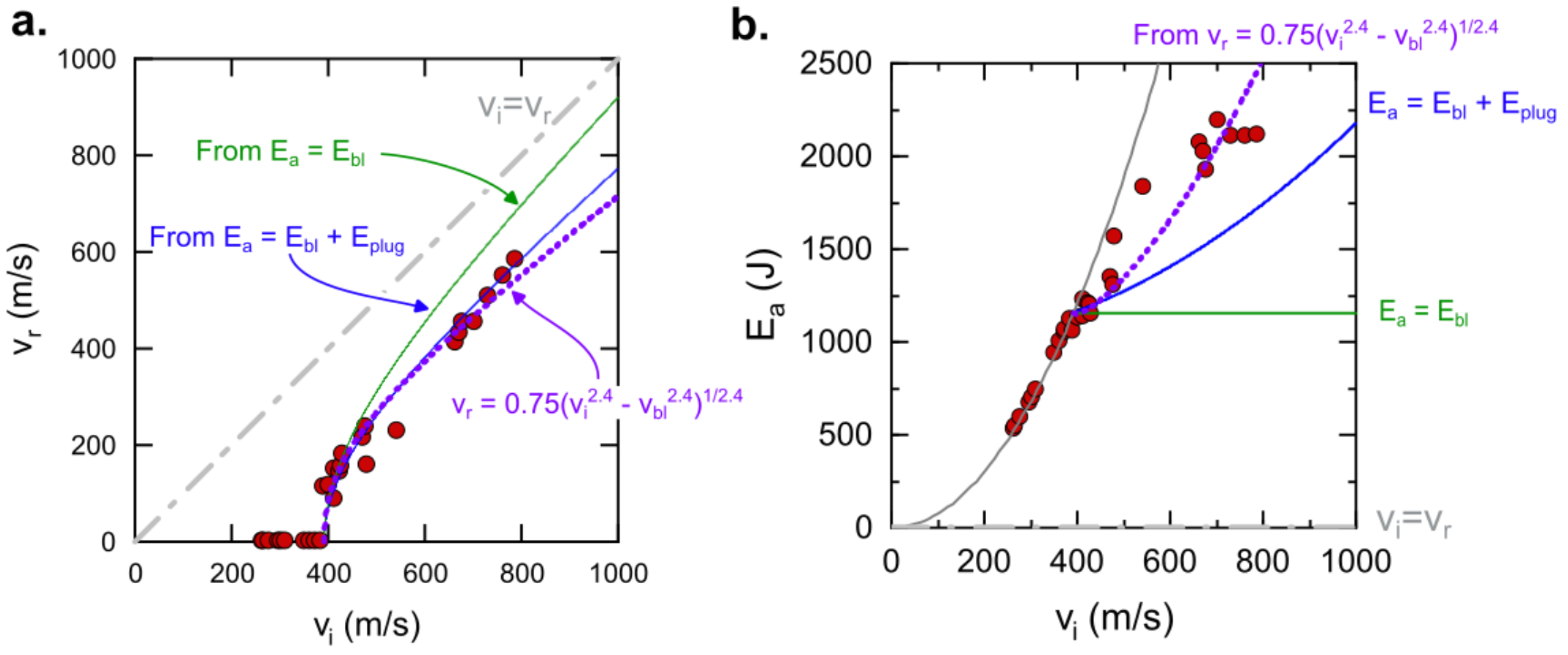

**Suppl. Fig. 1: Impact response prediction by popular models illustrating their limitations in describing experimental observations: a)** Residual velocity profile and **b)** Energy absorption as functions of the impact velocity. Red markers are experimental data corresponding to impacts performed on a $D/h \approx 3$ configuration. Green curves represent the Recht-Ipson model for a thin film where no plug mass exits, blue curves represent the Recht and Ipson model with plug mass. The dashed purple curves represent Lambert-Jonas fitting.

The Recht–Ipson model[1] was developed to provide a simplified description of residual-velocity trends. To achieve this, Recht–Ipson makes two central assumptions:

1. The projectile–plug system is treated as closed, allowing conservation laws to be applied directly.
2. The energy dissipated beyond the ballistic limit is assumed constant, regardless of the increasing impact velocity.

These assumptions yield a compact analytic expression for the residual-velocity curve. Two common variations are used in practice:

For thin targets with negligible plug mass

$$v_r = \sqrt{v_i^2 - v_{bl}^2}$$

For targets where a plug mass exits the plate

$$v_r = \frac{m_p}{m_p + m_{\text{plug}}}\sqrt{v_i^2 - v_{bl}^2}$$

The corresponding predictions are shown by the green and blue curves in Supplementary Fig. 1.

In both forms, the Recht–Ipson model predicts the residual velocity, but the corresponding energy dissipation is not independently predicted—it is prescribed through the assumption that the work required for perforation remains constant beyond the ballistic limit. The resulting $v_i - v_r$ relation therefore follows directly from this assumed energy balance. Consequently, the model cannot be used to independently infer how energy absorption evolves with increasing impact velocity, since doing so would recover a behavior already imposed in the formulation.

A second important assumption concerns the momentum balance. The formulation treats the projectile and the ejected plug as the interacting system for momentum conservation. Physically, however, the projectile and plug do not constitute a closed system because the plug represents only part of the responding target. Stress waves propagate through the surrounding target and carry momentum away from the projectile–plug region, a contribution not captured by this momentum balance. The formulation therefore effectively assumes that the momentum transferred to the remainder of the target is negligible. Notably, Recht and Ipson later introduced an impulse-measurement method using a momentum trap to quantify the momentum transferred to the surrounding target, demonstrating that this contribution need not be insignificant[2]. This further highlights the limitations of treating the projectile–plug subsystem as closed.

A related fitting-based description is the Lambert–Jonas model[3], which generalizes the Recht–Ipson residual-velocity relation as

$$v_r = a\,(\,v_i^p - v_{bl}^p)^{1/p}$$

Here $a$ and $p$ are obtained by fitting experimental ballistic data. For the present fit, a= $0.75$ and $p = 2.4$, and these fitting coefficients do not represent specific physical properties or mechanisms of the projectile–target interaction. The model can therefore reproduce the measured residual-velocity trend within the fitted range, but its extrapolation does not necessarily remain physically meaningful. In particular, if a fit yields $a > 1$, the model asymptotically approaches $\frac{v_r}{v_i} \to a > 1$, resulting in the nonphysical prediction $v_r > v_i$ at sufficiently high impact velocities.

## Supplementary Note 2: Summary of models for ballistic perforation

**Supplementary Table 1. Representative models used to describe ballistic penetration and perforation.**

| Model / study | Model basis / perforation criterion |
|---|---|
| *Energy- and work-based descriptions* | |
| **Recht–Ipson (1963)**[1] | Inelastic projectile–plug momentum balance; additional work associated with the peripheral shear resistance is determined at the ballistic limit and treated as approximately velocity-insensitive. |
| **Morye et al. (2000)**[4] | Ballistic-limit energy balance; projectile kinetic energy is absorbed through tensile failure, elastic deformation, and kinetic energy of the moving target cone. |
| **Naik et al. (2006)**[5] | Stress-wave propagation defines the evolving target region; incremental energy balance and material failure strains govern shear plugging, tensile failure, and other energy-absorbing mechanisms. |
| **López-Puente et al. (2007)**[6] | Differential energy balance; projectile kinetic-energy loss is consumed by laminate crushing, acceleration of detached target material, and tensile-fiber failure. |
| **Gama–Gillespie (2008)**[7] | Quasi-static punch-shear force–displacement response used to represent successive stages of ballistic penetration; accumulated penetration work characterizes target resistance. |
| **Cohen–Masri–Durban (2010)**[8] | Specific cavitation energy obtained from constitutive cavity-expansion fields; perforation occurs when projectile kinetic energy is sufficient to create the required hole volume. |
| *Structural-dynamics and failure-based descriptions* | |
| **Woodward (1987)**[9] | Linear and angular momentum balances determine the coupled projectile–plug and plate motion; perforation occurs through shear-plug formation or when membrane deformation reaches the tensile-fracture criterion. |
| *Force- and resistance-based descriptions* | |
| **Forrestal et al. (1990)**[10] | Cylindrical cavity expansion determines the target resistance from material strength and inertia; the resulting projectile force is integrated through the target to determine ballistic limit and residual velocity. |
| **Chen–Li (2003)**[11] | Cavity-expansion resistance governs the penetration stage; shear plugging initiates when projectile-nose resistance reaches the fully plastic shear resistance of the remaining target. |
| *Fitting-based models* | |
| **Lambert–Jonas (1976)**[3] | Generalized incident–residual velocity relation fitted to ballistic-test data; (a), (p), and (v_{bl}) are obtained by nonlinear regression: $v_r = a\left(v_i^p - v_{bl}^p\right)^{1/p}$. |
| **THOR (1961)**[12] | Empirical regression of ballistic-test data; residual velocity and residual projectile mass are represented as functions of projectile and target parameters. |

**Supplementary Note 3: Average chain length of polystyrene**
The average end-to-end chain length is calculated as:

$$R_0 = \sqrt{< R^2 >} = \sqrt{nC_\infty l^2} \quad (1)$$

Substituting $n = 2N$, where $N = \frac{280\ 000}{104.15}$ is the number of monomer units, $C_\infty$=9.6, and $l = 0.154\ nm$ for C-C bond length for polystyrene[13], we obtain the average end-to-end chain length for polystyrene as $R_0 = 35\ nm$.

**Supplementary Note 4: Polystyrene target fabrication parameters**
Suppl. Table 2 presents the polystyrene-to-toluene concentrations and the spin coater RPMs that were employed to fabricate the different thickness targets

**Supplementary Table 2. PS-toluene concentration (wt.%), spin-coating RPM, and PS film thickness (h)**

| $D/h$ | $D = 3.2\ \mu m$ | | | $D = 8.5\ \mu m$ | | | $D = 22.4\ \mu m$ | | | $D = 2\ mm$ | | |
|---|---|---|---|---|---|---|---|---|---|---|---|---|
| | wt.% | RPM | $h\ (\mu m)$ | wt.% | RPM | $h\ (\mu m)$ | wt.% | RPM | $h\ (\mu m)$ | wt.% | RPM | $h\ (mm)$ |
| 3 | 6 | 500 | 0.99 | 8 | 500 | 2.80 | 18 | 1500 | 6.900 | | | |
| 6 | 4 | 500 | 0.57 | 10 | 1650 | 1.47 | 15 | 1500 | 3.609 | | | |
| 10 | 4 | 3000 | 0.33 | 6 | 600 | 0.889 | 8 | 900 | 2.130 | 30 | 150-250 | 0.196-0.233 |

**Supplementary Note 5: Velocity measurement and air-drag correction**

The incident and residual velocities of the projectile are estimated by tracking the evolution of the projectile positions captured via the multi-exposure camera. The white light pulsation timescale is infinitesimal—in the order of 100s of nanoseconds–attributing a high sensitivity for velocity measurements on the projectile position. The error bars in Suppl. Fig.2b show the velocity variation ($\pm 6\ m/s$) when the center position is moved by a single pixel, attributing a significant human error if the centers are selected manually. To address this challenge and to account for the deceleration of the projectile due to air-drag, we developed the following velocity measurement technique.

The multi-exposure camera captures the side-view of the impact event (see Manuscript Fig.1b(iv)), where the dark circles define the temporal evolution of the microparticle positions recorded at each white-light laser pulse at a time interval $\Delta t$. First, the particle center positions, $Z_p = z_p^0, z_p^1, z_p^2, \dots z_p^n$, are determined manually and the average velocities are calculated with successive positions using:

$$v^k = \frac{z_p^{k+1} - z_p^k}{\Delta t} \quad (2)$$

Accuracy of the particle center positions depend on the pixelated camera snapshot, as enlarged in the inset of Suppl. Fig.2a, that propagates uncertainties in the velocity measurement. Hence, the center positions are perturbed iteratively such that the average deceleration profile is uniform:

$$a^k = \frac{v^{k+1} - v^k}{\Delta t}. \quad (3)$$

Since manual detection of the projectile center positions can incur significant human-errors and the projectile decelerates non-linearly due to air-drag, a MATLAB script was employed to further perturb each of the center positions such that the velocity profile agrees with the following air-drag corrected model. Assuming perfect spherical projectiles with negligible surface roughness, the deceleration is calculated as:

$$m_p\left(\frac{dv}{dt}\right) = -\frac{1}{2}C_D\rho_{air}Av^2\ , \tag{4}$$

where $m_p$ is the microprojectile mass, $C_D$ is the drag coefficient, $\rho_{air}$ is the air density, $A$ is the cross-sectional area of the microparticle, and $v$ is the microparticle velocity. Solving the above partial differential equation and reintegrating it provides closed-form equations for the particle velocity and position:

$$v(t) = \frac{v_0}{Bv_0(t-t_0)+1} \tag{5}$$

$$z(t) = \frac{\left(t-t_0+\left(\frac{1}{Bv_0}\right)-ln\left(\frac{1}{Bv_0}\right)\right)}{B}+z_0 \tag{6}$$

where $B=\frac{C_D\rho_{air}A}{2m_p}$, $v_0$ and $z_0$ are the average velocity and position at $t_0$. Manually measured incident $Z_p$ along with the corresponding times $(0, \Delta t, 2\Delta t, \ldots n\Delta t)$ are fitted to Eq.(6) where $v_0$ is obtained as the fitting parameter. The time of impact $(t_i)$ is estimated using the fitted model by taking the corresponding positions: $z(t_i) = z_0 - \frac{D}{2}$, where $z_0$ is the film top position. The impact velocity is extracted using Eq.(5) substituting for $t = t_i$. The same process is repeated for the residual velocity where $z(t_r) = z_0 + \frac{D}{2}$. The difference between the manually extracted positions $Z_p$ and the fitted model $z(t)$ are found to be within 1 μm across all measurements, that correspond to sub-pixel position corrections, independently verifying the accuracy of the air-drag model and the accurate extraction of projectile centers.

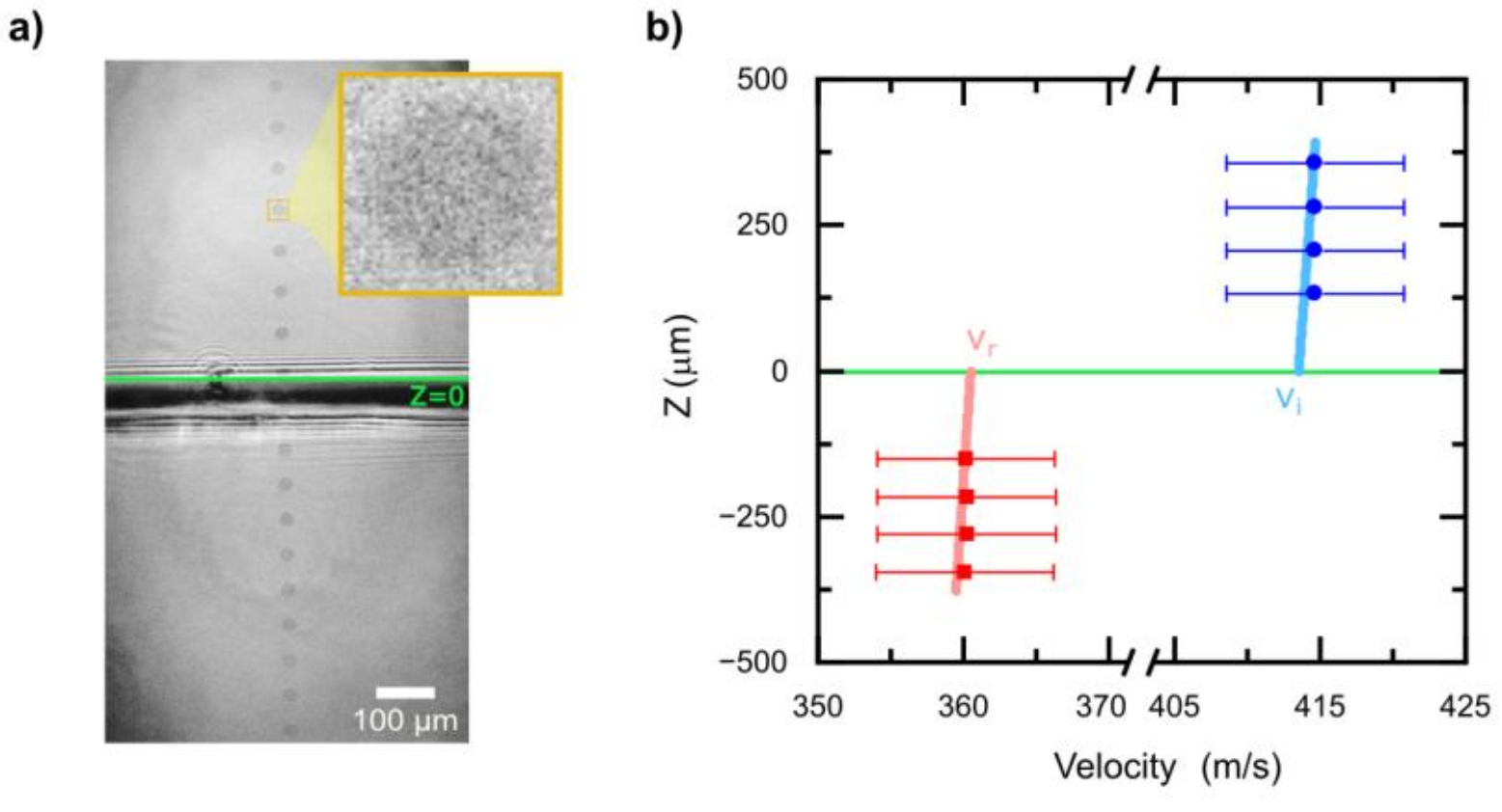

**Suppl. Fig. 2: Projectile's velocity measurement using the multi-exposure camera. (a)** The projectile path corresponding to each white-laser light pulse at equal time intervals. **(b)** The estimated incident (blue) and residual (red) velocities calculated by the air-drag correction MATLAB code. The error bars correspond to velocity variations if the projectile position is perturbed by a single pixel.

**Supplementary Note 6: Uncertainty quantification for momentum transfer and energy transfer measurements**

The supplier-specified nominal diameters were 4.08, 9.20, and 20 µm. However, SEM measurements of 662, 152, and 208 particles, respectively, yielded mean diameters of 3.2, 8.5, and 22.4 µm, with relative standard deviations of 7.5%, 4.8%, and 6.3% (Suppl.Fig.3).

The accurate calculation of momentum and kinetic energy relies on the certainty of two primary quantities: the projectile's velocity and mass. Here, we briefly discuss the potential sources of errors, and the detailed variations are reported in Supplementary Table 3.

*Mass measurement*

The projectile mass, $m_p = \frac{1}{6}\rho_p \pi D^3$, depends on the projectile diameter and the density. We use the supplier data sheet for the projectile density, and the diameter distributions are measured via SEM imaging. The projectile sizes have a 5-7% variation which translates to a 15-22% variation in the mass. Note that this uncertainty does not account for potential density variations or surface roughness of the projectile.

*Velocity measurements*

The impact velocity of a microparticle is measured via the two step-process described in Supplementary Note 5: first the manual identification of microparticle centers and then automated perturbation of the centers based on the air-drag correction model. The air-drag correction is based on the projectile mass, see Eq. (4), therefore mass measurement uncertainties need to be propagated towards the velocity estimation. Using the extreme diameter values from the measured variations: $D_{avg} - \sigma_D$ and $D_{avg} + \sigma_D$, air-drag corrected velocities are recalculated and the resulting velocity deviations are found to be less than 1%.

**Supplementary Table 3. Uncertainty propagation**

| Mean projectile size (µm) | 3.2 | 8.5 | 22.4 |
|---|---|---|---|
| Diameter, $\frac{\sigma_D}{D}$ | 7.5 % | 4.8% | 6.3% |
| Projectile mass, $\frac{\sigma_{mp}}{m_p}$ | 22.5 % | 14.4 % | 18.9 % |
| Velocity, $\frac{\sigma_v}{v}$ | 0.6 % | 0.15 % | 0.03 % |
| Normalized momentum, $\frac{\sigma_{\Delta\tilde{P}}}{\Delta\tilde{P}}$ | 31.8 % | 20.4% | 26.7 % |
| Normalized energy, $\frac{\sigma_{\tilde{E}_a}}{\tilde{E}_a}$ | 31.8 % | 20.4 % | 26.7 % |

Note that velocity measurement through the air-drag correction model is subjected to assumptions of smooth spheres with drag coefficients predicted by the $C_d$ curve. Any errors arising from the model will be systematic errors, affecting both the incident and residual velocities in a similar manner. If these systematic errors are identified in future, they can be corrected.

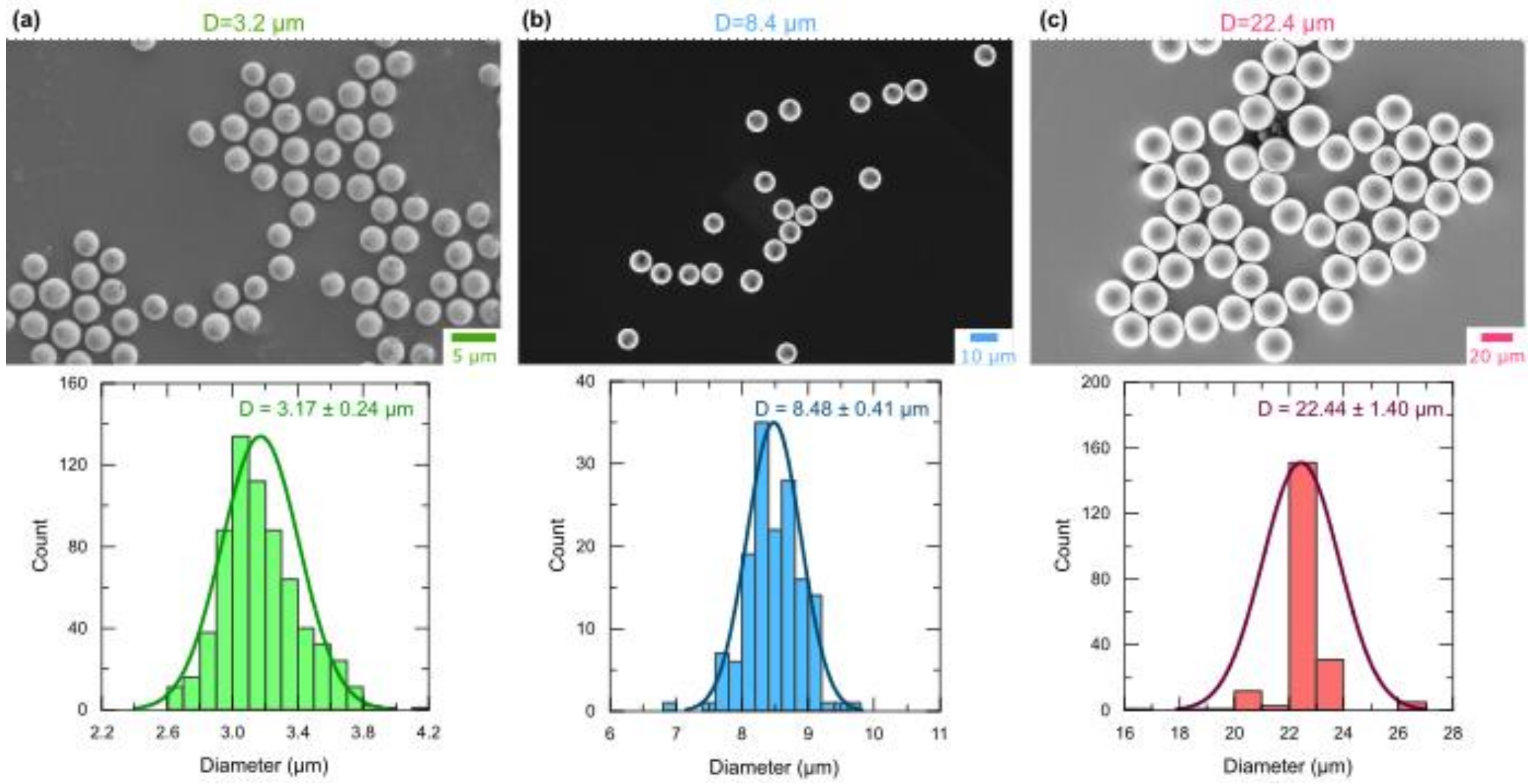

**Suppl. Fig. 3: Projectile size distribution measured via SEM imaging, and the normal distribution fitting parameters. (a-c)** Example SEM images for 3.2 µm, 8.5 µm, and 22.4 µm particles and the statistical distributions for total number of particles: 662, 152, and 208, respectively.

## Supplementary Note 7: Energy absorption and Specific Energy absorption plots

The complete set of energy absorption and Specific Energy absorption data, including the arrested data, are shown in Suppl.Fig. 4.

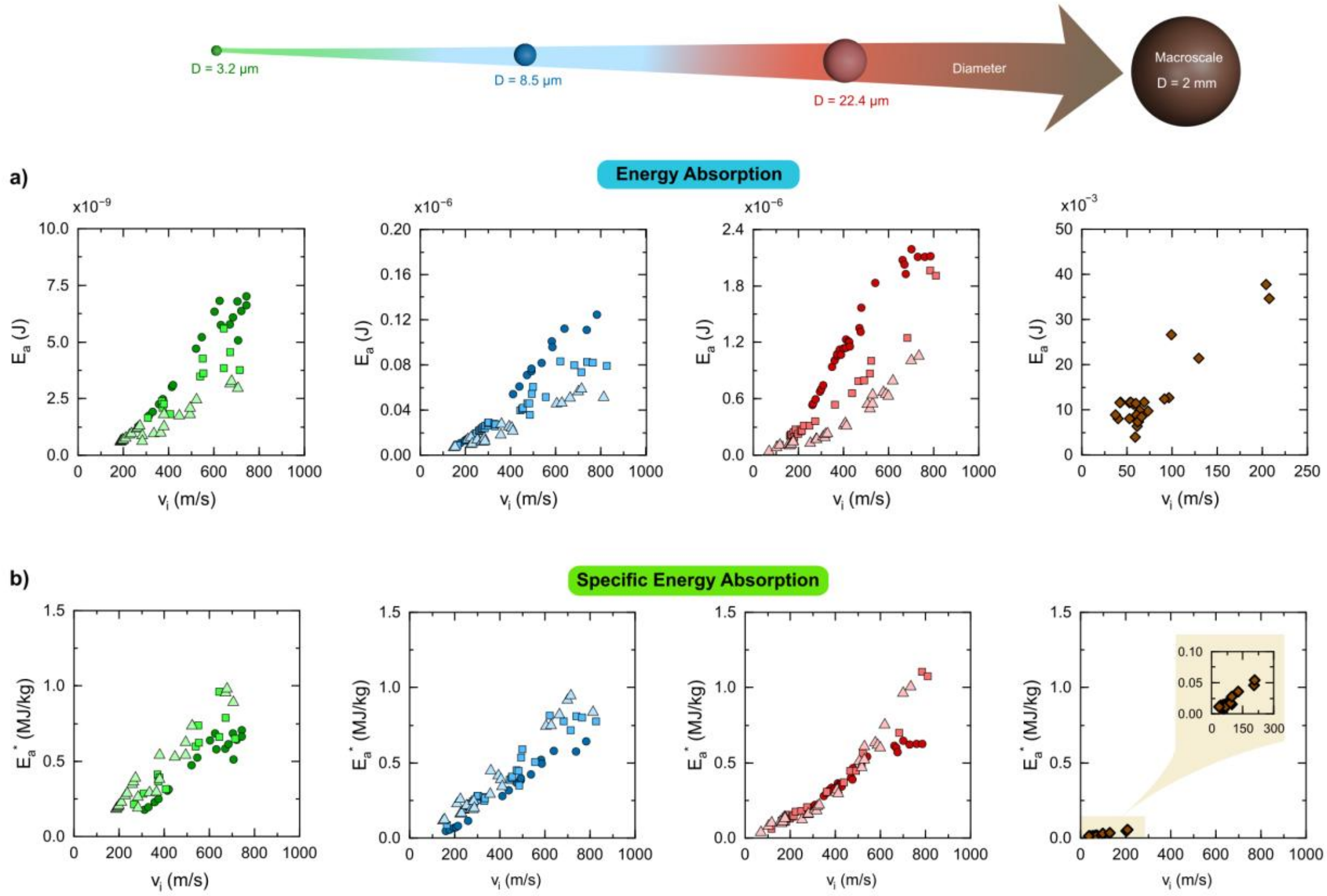

**Suppl. Fig. 4:** All **(a)** Energy absorption and **(b)** Specific energy absorption values as a function of the impact velocity for each projectile-target combination. The marker shapes and colors follow Fig.1c where circle, square, and triangle shapes are represented $D/h \approx 3, 6$, and 10, respectively. Macro scale results are also included.

**Supplementary Note 8: Extending the momentum transfer bound to the energy landscape**

We observe the momentum transfer upon impact to be bounded by the momentum transfer at the ballistic limit, $\Delta\tilde{P} < 1$, which can be simplified as:

$$\Delta\tilde{P} = \frac{m_p(v_i - v_r)}{m_p v_{bl}} < 1 \tag{7}$$

$$v_r > v_i - v_{bl} \tag{8}$$

Substituting this relation on the normalized energy absorption,

$$\tilde{E}_a = \frac{\frac{1}{2} m_p (v_i^2 - v_r^2)}{\frac{1}{2} m_p v_{bl}^2} \tag{9}$$

$$\tilde{E}_a < \frac{v_i^2 - (v_i - v_{bl})^2}{v_{bl}^2} \tag{10}$$

$$\tilde{E}_a < \frac{(2v_i - v_{bl})v_{bl}}{v_{bl}^2} \tag{11}$$

$$\boldsymbol{\tilde{E}_a < 2\tilde{v}_i - 1} \tag{12}$$

Extending this condition to the dimensional case:

$$E_a < E_{bl}(2\tilde{v}_i - 1) \tag{13}$$

$$E_a < \frac{1}{2} m_p v_{bl}^2 \frac{2v_i}{v_{bl}} - E_{bl} \tag{14}$$

$$E_a < m_p v_{bl} v_i - E_{bl} \tag{15}$$

$$\boldsymbol{E_a < \Delta P_{bl} v_i - E_{bl}} \tag{16}$$

Dividing by the ideal plug mass:

$$\frac{E_a}{m_{plug}} < \frac{\Delta P_{bl}}{m_{plug}} v_i - \frac{E_{bl}}{m_{plug}} \tag{17}$$

$$\boldsymbol{E_a^* < \Delta P_{bl}^* v_i - E_{bl}^*} \tag{18}$$

## Supplementary Note 9: Estimating the nominal strain rate, penetration time, and Region of influence

While most projectile impact studies define the nominal strain rate as $\dot{\varepsilon}_{nom} = \frac{v_i}{D}$, this definition imposes the same strain rate irrespective of the target thickness. For example, same $v_i$ and $D$ on a thicker sample might arrest the projectile, while a thinner target might be perforated with infinitesimal deceleration. It would be erroneous to associate both cases with the same strain rate, therefore, we use $v_{avg}$ instead of $v_i$. Here, the rationale is that $v_r$ would vary based on the target thickness.

$$\dot{\varepsilon}_{nom} = \frac{v_{avg}}{D}, \tag{19}$$

where,

$$v_{avg} = \frac{(v_i + v_r)}{2}. \tag{20}$$

The evolution of $v_{avg}$ and $\dot{\varepsilon}_{nom}$ are shown in Suppl. Fig.5a and b for all the impact geometries, which increase with $v_i$.

The minimum penetration time corresponds to the time the projectile takes to traverse the thickness of the polystyrene target,

$$t_{min} = \frac{h}{v_{avg}}, \tag{21}$$

which steadily decrease with increasing $v_i$. The SEM images indicate that the polystyrene targets undergo much larger stretching before failure. However, the minimum perforation time provides a qualitative understanding of the impulse duration with increasing impact velocity (see Suppl. Fig.5c).

The ROI radius is calculated as:

$$ROI = c_0\, t_{min}, \tag{22}$$

where $c_0 = \sqrt{\frac{E_t}{\rho_t}}$ is the elastic wave speed of the target (see Suppl. Fig.5d). Although these ROI estimates correspond to the minimum penetration time, they illustrate the localization phenomena with increased velocity.

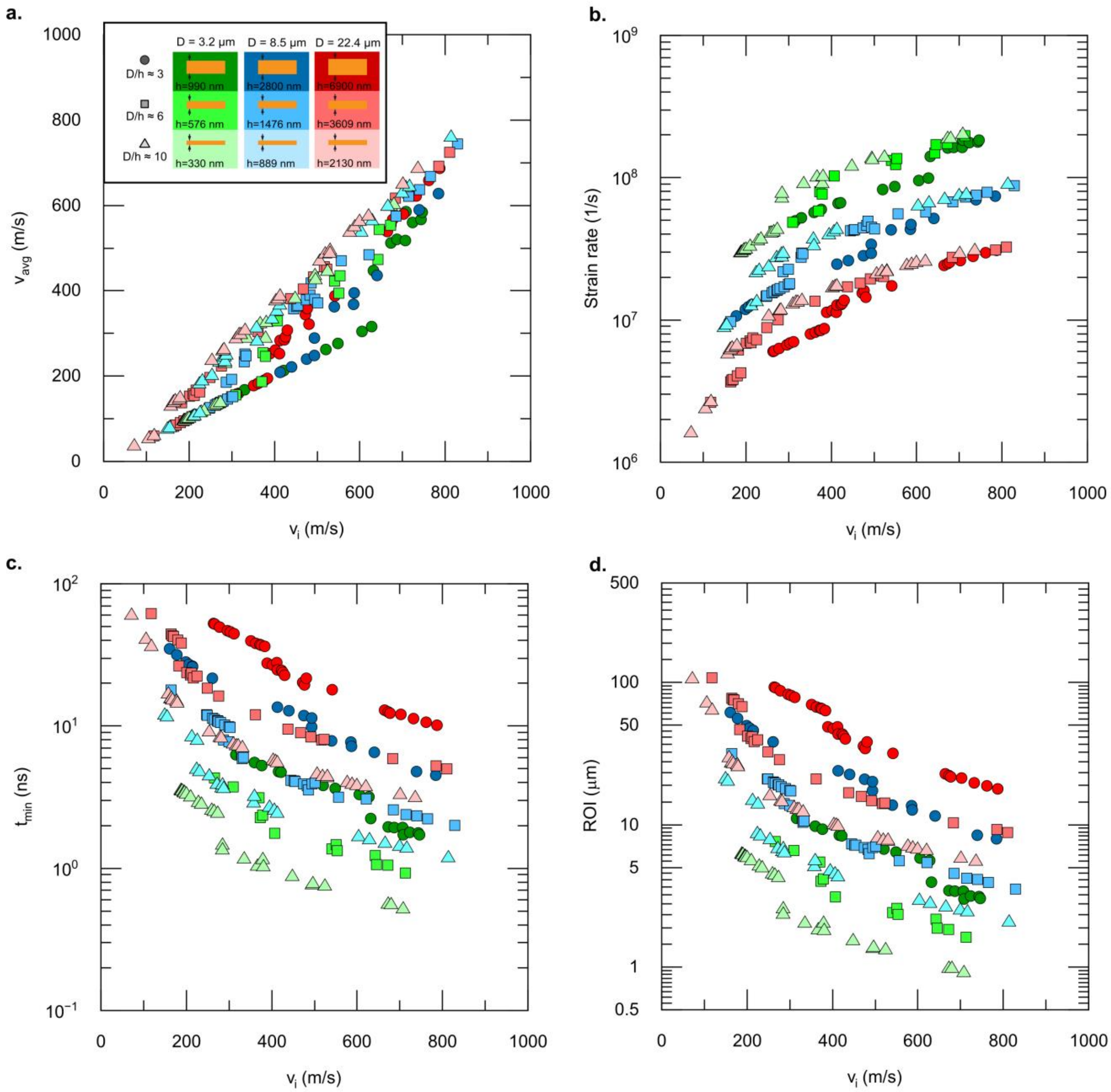

**Suppl. Fig.5**: Estimation of (a) $v_{avg}$, (b) average strain rate, (c) minimum penetration time, and (d) radius of region of influence as a function of $v_i$ for all the geometries.

## Supplementary Note 10: Calculation of the minimum inertial momentum transfer

The post-mortem SEM illustrates that the ROI deforms upon contact, transferring momentum to accelerate the target mass. Hence, we establish the condition that the ideal plug mass, $m_{plug} = \frac{\pi \rho_t D^2 h}{4}$, obtains a minimum instantaneous velocity of $v_r$ for the projectile for target penetration. The change in the projectile's momentum to accelerate $m_{plug}$ is:

$$\Delta P_{min} = m_p(v_i - v_r) = m_{plug} v_r, \tag{23}$$

which leads to the relation:

$$v_r = \frac{m_p}{m_{plug}+m_p} v_i. \tag{24}$$

Hence, the transfer of projectile's momentum as a function of $v_i$:

$$\Delta P_{min} = \frac{m_{plug} m_p}{m_{plug} + m_p} v_i \tag{25}$$

Normalizing by $\Delta P_{bl}$ yields:

$$\Delta \tilde{P}_{min} = \frac{\Delta P_{min}}{\Delta P_{bl}} \tag{26}$$

$$\Delta \tilde{P}_{min} = \left( \frac{m_{plug} m_p}{m_{plug} + m_p} v_i \right) \frac{1}{m_p v_{bl}} \tag{27}$$

$$\Delta \tilde{P}_{min} = \frac{m_{plug}}{m_{plug}+m_p} \tilde{v}_i = \zeta \tilde{v}_i \tag{28}$$

where $\zeta = \frac{m_{plug}}{m_{plug}+m_p} < 1$ is a nondimensional mass ratio between the projectile and the ideal plug mass. Ideally, $\zeta \sim \frac{1}{1+D/h}$ depends only on the $D/h$ ratio, however, small variations are observed in Supplementary Table 4 due to different projectile densities and fabricated thicknesses only approximate the three different geometric ratios.

Equation (28) shows that the inertial contribution of the momentum transfer increases with incident velocity, and can even exceed the experimentally observed momentum transfer bound, $\Delta \tilde{P}_{min} > 1$, when $v_i > \frac{v_{bl}}{\zeta}$. Most ballistics tests do not observe this phenomenon as the plug mass is much smaller than the projectile mass ($m_{plug} < m_p$), therefore, $\zeta < 1$ and $\frac{v_{bl}}{\zeta}$ is much larger than the typically tested ballistic velocity ranges. However, it is important to recognize the limitations of the observed bounds.

**Supplementary Table 4. $\zeta$ values for the minimum momentum transfer**

| $\zeta$ | $D = 3.2\ \mu m$ | $D = 8.5\ \mu m$ | $D = 22.4\ \mu m$ |
|---|---|---|---|
| $D/h \approx 3$ | 0.1976 | 0.2089 | 0.1593 |
| $D/h \approx 6$ | 0.1252 | 0.1209 | 0.0902 |
| $D/h \approx 10$ | 0.0758 | 0.0765 | 0.0553 |

The minimum momentum transfer in the normalized energy space is obtained using Equation (24) for $v_r$, which yields:

$$\tilde{E}_{a,min} = \frac{v_i^2 - v_r^2}{v_{bl}^2} = \zeta(2-\zeta)\tilde{v}_i^2. \quad (29)$$

and the coefficients for each geometry are shown in Supplementary Table 5.

**Supplementary Table 5. $\zeta(2-\zeta)$ coefficients for minimum energy transfer**

| $\zeta(2-\zeta)$ | $D = 3.2\ \mu m$ | $D = 8.5\ \mu m$ | $D = 22.4\ \mu m$ |
|---|---|---|---|
| $D/h \approx 3$ | 0.3561 | 0.3738 | 0.2933 |
| $D/h \approx 6$ | 0.2347 | 0.2272 | 0.1722 |
| $D/h \approx 10$ | 0.1458 | 0.1472 | 0.1075 |

**Supplementary Note 11: Reducing momentum transfer trend when energy transfer saturates**

When the kinetic energy transfer saturates at,

$$E_a = \frac{1}{2} m_p (v_i^2 - v_r^2) = E_{sat} \,, \tag{30}$$

residual velocity is expressed as

$$v_r = \sqrt{v_i^2 - \frac{2E_{sat}}{m_p}} \tag{31}$$

$$v_r = v_i \sqrt{1 - \Omega} \tag{32}$$

where $\Omega = \frac{2E_{sat}}{m_p v_i^2} = \frac{E_{sat}}{E_i}$. The momentum transfer is:

$$\Delta P = m_p (v_i - v_i \sqrt{1 - \Omega}) \tag{33}$$

$$\Delta P = m_p v_i \left(1 - \sqrt{1 - \Omega}\,\right) \tag{34}$$

Since $|\Omega| < 1$, we apply the binomial expansion on $\sqrt{1 - \Omega} = 1 - 0.5\Omega - 0.125\Omega^2 - \cdots$

$$\Delta P = m_p v_i (1 - (1 - 0.5\Omega - h.o.t)\,) \tag{35}$$

Neglecting the higher order terms,

$$\Delta P \approx m_p v_i \left(\frac{1}{2}\Omega\right) \tag{36}$$

$$\Delta P \approx m_p v_i \left(\frac{1}{2} \frac{2E_{sat}}{m_p v_i^2}\right) \tag{37}$$

$$\Delta P \approx \frac{E_{sat}}{v_i} \tag{38}$$

Hence, when kinetic energy transference is saturated, increasing impact velocities result in reducing momentum change.

**Supplementary Note 12: Contradictory ranking of ballistic performance using specific energy absorption and specific momentum transfer**

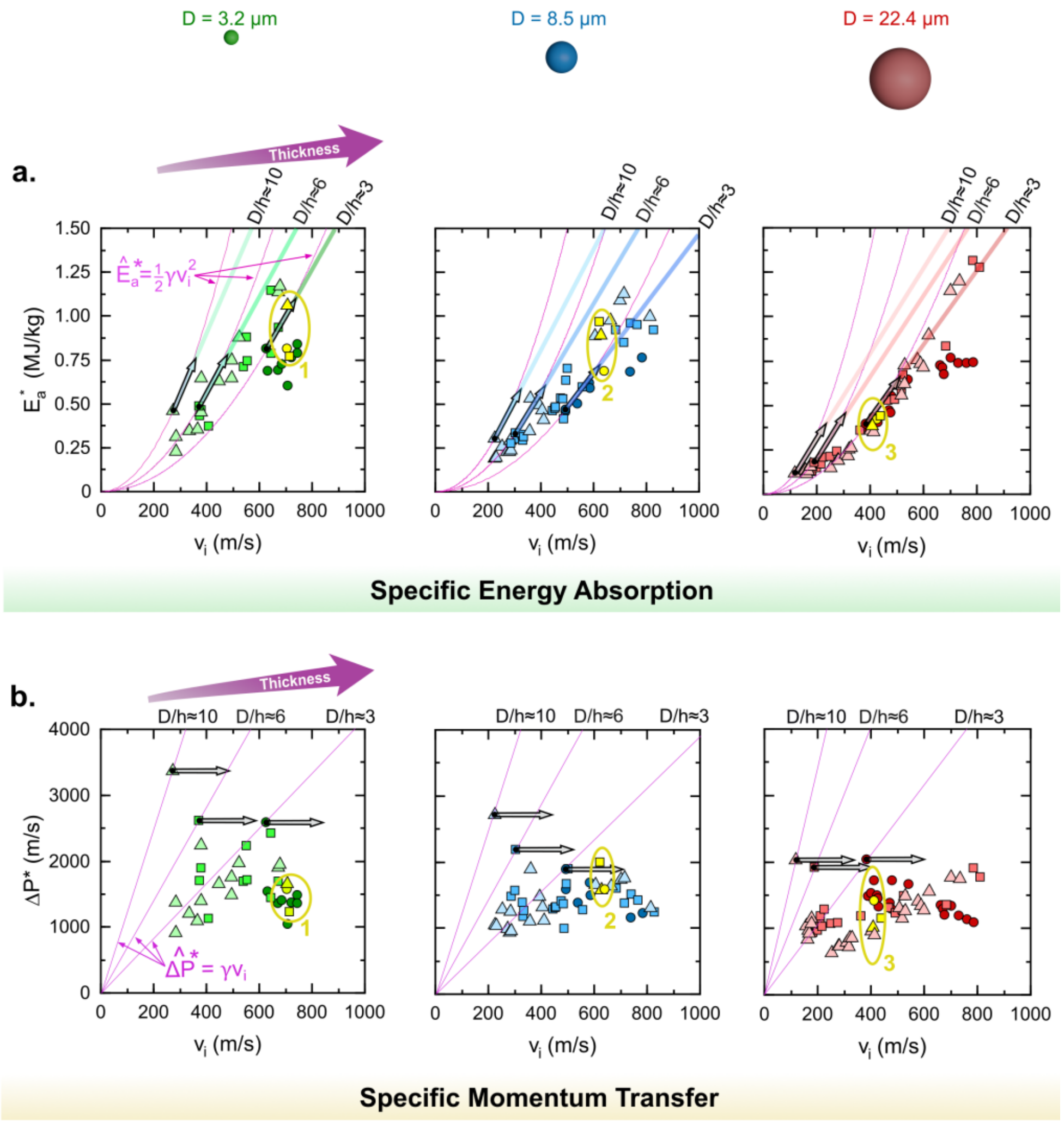

**Suppl. Fig.6**: Specific energy absorption and specific momentum transfer plots highlighting the ranking conclusions. The same points are marked in yellow for easy recognition.

To further illustrate the inconsistency between specific energy absorption, and specific momentum transfer three representative comparisons are highlighted in Suppl. Fig. 6. The same data points are identified by yellow circles in both the specific-energy and specific-momentum landscapes. In each case, the ranking inferred from $E_a^*$ differs from that inferred from $\Delta P^*$, demonstrating that the choice of metric alone can alter the apparent ordering of ballistic performance.

These examples demonstrate that specific energy absorption and specific momentum transfer do not simply rescale the same information. Rather, they can lead to qualitatively different conclusions regarding the relative ballistic performance of target configurations. In Circle 1, the specific-energy metric suggests that the 330 nm target substantially outperforms both thicker targets, whereas the specific-momentum metric indicates that the 330 nm and 576 nm targets perform equivalently. In Circle 2, the specific-energy metric identifies two top-performing targets, whereas the specific-momentum metric identifies a single best-performing target. In Circle 3, the specific-energy metric suggests all three targets perform similarly, while the specific-momentum metric reveals a clear performance hierarchy.

**Supplementary Table 6. Summary of specific energy absorption and specific momentum transfer rankings in Suppl. Fig.6**

| Case | Interpretation based on specific energy absorption, $E_a^*$ | Interpretation based on specific momentum transfer, $\Delta P^*$ |
|---|---|---|
| Circle 1 | The 330 nm target is identified as the highest-performing configuration, with $E_a^* \approx$ 1.0MJ/kg. The 576 nm and 990 nm targets appear comparable at $E_a^* \approx$ 0.75MJ/kg. | The 330 nm and 576 nm targets exhibit equivalent performance, both reaching $\Delta P^* \approx$ 1700m/s. The 990 nm target is lower, with $\Delta P^* \approx$ 1200m/s. |
| Circle 2 | The 889 nm and 1476 nm targets are identified as equivalent high-performing configurations, with $E_a^* \approx$ 0.8MJ/kg. The 2800 nm target appears lower, with $E_a^* \approx$ 0.7MJ/kg. | The 1476 nm target is identified as the highest-performing configuration, with $\Delta P^* \approx$ 2000m/s. The 889 nm and 2800 nm targets exhibit similar performance, with $\Delta P^* \approx$ 1500m/s. |
| Circle 3 | The 2130 nm, 3609 nm, and 6900 nm targets appear to exhibit comparable performance, all with $E_a^* \approx$ 0.4MJ/kg. | A clear performance hierarchy emerges: the 6900 nm target exhibits the highest performance, with $\Delta P^* \approx$ 1500m/s, followed by the 3609 nm target with $\Delta P^* \approx$ 1100m/s, and the 2130 nm target with $\Delta P^* \approx$ 1000m/s. |

**Supplementary Note 13: Multilayered target impact response**

Consider a multilayered target having $n$ layers where the individual layer thickness is $h$ and ballistic limit velocity $v_{bl}$. Let the projectile's incident velocity to the top layer is $v_1 = v_i$, and the subsequent incident velocity to each layer to $v_2, v_3, \ldots. v_n$, with the final residual velocity $v_{n+1} = v_r$. For the $k^{th}$ layer, the momentum transfer upper bound in energy absorption terms:

$$E_{a,k} = \Delta P_{bl} v_k - E_{bl} \tag{39}$$

The total energy absorption of the multilayered target is:

$$E_{a,Total} = \Sigma_{k=1}^{n} \left(E_{a,k}\right) \tag{40}$$

$$E_{a,Total} = \Delta P_{bl}\, \Sigma_{k=1}^{n} (v_k) - E_{bl} \Sigma_{k=1}^{n} (1) \tag{41}$$

Using the condition $v_r = v_i - v_{bl}$ from momentum transfer bound (see Eq: (8)) and applying between two successive layers iteratively yields: $v_k = v_{k-1} - v_{bl} = v_{k-2} - 2v_{bl} = \cdots = v_1 - (k-1)v_{bl}$.

$$E_{a,Total} = \Delta P_{bl}\, \Sigma_{k=1}^{n} (v_i - (k-1)v_{bl}) - E_{bl} \Sigma_{k=1}^{n} (1) \tag{42}$$

$$E_{a,Total} = \Delta P_{bl,B-single} \left( \mathrm{n} v_1 - \frac{n(n+1)}{2} v_{bl} + n v_{bl} \right) - n E_{bl} \tag{43}$$

$$E_{a,Total} = m_p v_{bl} \left( \mathrm{n} v_1 - \frac{n(n+1)}{2} v_{bl} + n v_{bl} - \frac{n}{2} v_{bl} \right) \tag{44}$$

$$E_{a,Total} = n m_p v_{bl} \left( v_1 - \frac{1}{2} n v_{bl} \right) \tag{45}$$

$$E_{a,Total} = m_p (n v_{bl}) \left( v_i - \frac{1}{2} (n v_{bl}) \right) \tag{46}$$

$$E_{a,Total} = m_p (n v_{bl}) v_i - \frac{1}{2} m_p (n v_{bl})^2 \tag{47}$$

Substituting $n v_{bl} = v_{bl-Total}$ yields:

$$\boldsymbol{E_{a,Total} = \Delta P_{bl-Total} v_i - E_{bl-Total}} \tag{48}$$

Therefore, the momentum transfer bound for the multilayered target is defined by the tangent at $v_i = n v_{bl}$.

For specific energy absorption, total plug mass is $n m_{plug}$:

$$E^*_{a,Total} = \frac{E_{a,Total}}{n m_{plug}}$$

From Eq.(47),

$$E^*_{a,Total} = \frac{m_p}{n m_{plug}} (n v_{bl}) v_i - \frac{1}{2} \frac{m_p}{n m_{plug}} (n v_{bl})^2 \tag{49}$$

$$E^*_{a,Total} = \gamma (v_{bl}) v_i - \frac{1}{2} n \gamma\, (v_{bl})^2 \tag{50}$$

$$E^*_{a,Total} = \gamma(v_{bl})v_i - \frac{1}{2}\gamma\ (v_{bl})^2 - \frac{1}{2}(n-1)\gamma\ (v_{bl})^2 \quad (51)$$

$$\boldsymbol{E^*_{a,Total} = E^*_a - (n-1)E^*_{bl}} \quad (52)$$